\documentclass[useAMS,usenatbib]{mn2e}
\usepackage{times,txfonts,graphicx,aas_macros,natbib,xspace,color}
\def\icarus{{Icarus}}
\usepackage{ulem}

\renewcommand\emph[1]{\textit{#1}}

\newcommand\cm{\,\rm cm}
\newcommand\m{\,\rm m}
\newcommand\s{\,\rm s}
\newcommand\ms{\,\rm m\,s^{-1}}
\newcommand\g{\,\rm g}

\newcommand\K{\,\rm K}
\newcommand\yr{\,\rm yr}
\newcommand\Myr{\,\rm Myr}

\newcommand\km{\,\rm km}

\newcommand\au{\,\rm au}
\newcommand\mG{\,\rm mG}

\newcommand\perccm{\,\rm cm^{-3}}
\newcommand\Msun{\,\rm M_\odot}

\newcommand\tms{\!\times\!}
\newcommand\cdt{\!\cdot\!}

\newcommand\xx{\hat{{\mathbf x}}}
\newcommand\yy{\hat{{\mathbf y}}}
\newcommand\zz{\hat{{\mathbf z}}}

\newcommand\V{\mathbf v}
\newcommand\B{\mathbf B}

\newcommand\mB{\mn{\mathbf{B}}}

\newcommand\tauc{\tau_{\rm c}}

\newcommand{\mn}[1]{\overline{#1}}
\newcommand{\rms}[1]{\left<\right.\!#1\!\left.\right>}
\newcommand{\ee}[1]{$\,\times 10^{#1}$}

\newcommand\Rm{\mathrm{Rm}}
\newcommand\El{\Lambda}

\newcommand{\simgt}%
           {\,\hbox{\lower0.35ex\hbox{$\sim$}\llap{\raise0.35ex\hbox{$>$}}}\,}
\newcommand{\simlt}%
           {\,\hbox{\lower0.35ex\hbox{$\sim$}\llap{\raise0.35ex\hbox{$<$}}}\,}

\newcommand\NIII{\textsc{nirvana-iii}\xspace}

\title[Disc mass and field dependence of planetesimal stirring]%
      {Dead zones as safe-havens for planetesimals:
       influence of disc mass and external magnetic field}
      \author[Gressel, Nelson \& Turner]%
             { Oliver~Gressel$^{1\star}$,
               Richard~P.~Nelson$^{1\star}$ and
               Neal~J.~Turner$^2$\thanks{E-mail:~
               o.gressel@qmul.ac.uk (OG); r.p.nelson@qmul.ac.uk
               (RPN); neal.turner@jpl.nasa.gov (NJT)}\\
              $^1$Astronomy Unit, Queen Mary University of London,
               Mile End Road, London E1 4NS\\ 
               $^2$Jet Propulsion Laboratory, California Institute of
               Technology, Pasadena, CA 91109, USA}

\begin{document}

\date{Accepted 1988 December 15. %
      Received 1988 December 14; %
      in original form 1988 October 11}

\pagerange{\pageref{firstpage}--\pageref{lastpage}} \pubyear{2002}

\maketitle

\label{firstpage}


\begin{abstract}
Planetesimals embedded in a protoplanetary disc are stirred by
gravitational torques exerted by density fluctuations in the
surrounding turbulence. In particular, planetesimals in a disc
supporting fully developed magneto-rotational turbulence are readily
excited to velocity dispersions above the threshold for catastrophic
disruption, halting planet formation.
We aim to examine the stirring of planetesimals lying instead in a
magnetically-decoupled midplane dead zone, stirred only by spiral
density waves propagating out of the disc's magnetically-coupled
turbulent surface layers. We extend previous studies to include a
wider range of disc models, and explore the effects of varying the
disc column density and external magnetic field strength.
We measure the stochastic torques on swarms of test particles in 3D
resistive-MHD stratified shearing-box calculations with ionisation by
stellar X-rays, cosmic rays, and recombination on dust grains.
The strength of the stirring is found to be independent of the gas
surface density, which is contrary to the increase with disc mass
expected from a simple linear wave picture. The discrepancy arises
from the shearing out of density waves as they propagate into the dead
zone, resulting in density structures near the midplane that exert
weaker stochastic torques on average. We provide a simple analytic fit
to our numerically obtained torque amplitudes that accounts for this
effect. The stirring on the other hand depends sensitively on the net
vertical magnetic flux, up to a saturation level above which magnetic
forces dominate in the turbulent layers.
For the majority of our models, the equilibrium planetesimal velocity
dispersions lie between the thresholds for disrupting strong and weak
aggregates, suggesting that collision outcomes will depend on material
properties. However, discs with relatively weak magnetic fields yield
reduced stirring, and their dead zones provide safe-havens even for
the weakest planetesimals against collisional destruction.
\end{abstract}



\section{Introduction}
\label{sec:intro}

The formation of rocky earth-like planets, and the rock-ice cores of
gas giant planets, are believed to involve assembly from a population
of planetesimals that may range in size from 1 to $100\km$. The
existence of planets, asteroids and Kuiper-belt objects in our own
solar system covering a wide-range in semimajor axes, in addition to
the broad diversity in orbital properties of extrasolar planets
\citep[e.g.][]{1995Natur.378..355M,2010Natur.468.1080M,2011Natur.470...53L}
and debris discs around main-sequence stars
\citep{2008ARA&A..46..339W}, indicates that planetesimals form at
orbital distances ranging from a few tenths of an $\rm au$ out to more
than $100\,\au$ from the central star. While there remain many open
questions about how planetesimals form within a protoplanetary disc, a
number of plausible models and scenarios have been suggested that
exhibit differing degrees of sensitivity to the physical conditions
within the disc.

Classical models of planetesimal formation \citep[see
e.g.,][]{1993prpl.conf.1031W} involve an incremental process which
starts from micron-sized dust grains. These grains grow via mutual
collisions and sticking and settle toward the midplane of the
protoplanetary disc (PPD). Assuming a laminar nebula a few times
heavier than the minimum required to form the solar system
\citep{1981PThPS..70...35H}, \citet{2000SSRv...92..295W} predicts a
growth time scale equivalent to a few $\times 10^4\yr$ at $5\au$. The
main obstacle to the incremental formation scenario is the rapid
inward migration of growing boulders once they reach sizes around one
metre. Moreover, as demonstrated by \citet{2008A&A...480..859B},
differential radial migration of a population of planetesimals of
varying sizes can lead to high relative velocities implying
collisional fragmentation.

In an attempt to avoid these difficulties, alternative formation
scenarios have been proposed, which try to jump this metre-sized
barrier by involving mechanisms of rapid planetesimal formation. These
alternative scenarios go back to the classic idea of
\citet{1973ApJ...183.1051G}, who proposed formation of planetesimals
through gravitational instability in a dense layer of solids resulting
from vertical settling of dust grains. Alternative scenarios that
invoke rapid planetesimal formation include the concentration of
mm-sized chondrules in turbulent eddies followed by direct
gravitational collapse \citep{2008ApJ...687.1432C}, the streaming
instability \citep{2005ApJ...620..459Y}, and particle trapping in
zonal flows \citep{2009ApJ...697.1269J}. As demonstrated by
\citet{2007Natur.448.1022J,2011A&A...529A..62J}, processes of this
kind may ultimately lead to the rapid formation of bodies larger than
Ceres via gravitational collapse of local concentrations of
metre-sized objects, where the planetesimal masses depend on the disc
mass.

Owing to their size, the emergence of planetesimals marks the
transition into the gravitationally-dominated stage of planet
formation. Runaway growth quickly leads to the formation of oligarchs
\citep{1993Icar..106..190W,2009ApJ...690L.140K}, which subsequently
grow into planetary embryos and cores via mutual collisions and
accretion of smaller bodies \citep{1993Icar..106..210I,
1998Icar..131..171K}.  For this runaway growth to proceed efficiently,
the velocity dispersion within the planetesimal swarm must remain
significantly smaller than the surface-escape velocity of the
accreting cores. Accounting only for self-stirring of the population,
this is certainly expected in a fully laminar protostellar
nebula. However, the explanation of the typical accretion rates of
T~Tauri stars requires an anomalous source of viscous redistribution
of angular momentum, most likely explained by disc turbulence. This
poses the question whether or not runaway growth remains a realistic
proposition in a turbulent protoplanetary disc.

Planetesimals and protoplanetary embryos embedded in turbulent PPDs
are subject to stochastic gravitational forces caused by density waves
driven by the turbulent Maxwell stresses acting within the disc
\citep{2004MNRAS.350..849N}. This issue was addressed in our previous
work \citep[hereafter ``Paper~I'']{2010MNRAS.409..639N}, where we
compared global and local magneto-hydrodynamic (MHD) simulations of
protoplanetary discs involving turbulence driven by the
magneto-rotational instability
\citep[MRI,][]{1991ApJ...376..214B}. Following the evolution of swarms
of embedded test particles, we studied the effects of magnetic
turbulence on the dispersion of planetesimal semimajor axes and the
growth of their internal velocity dispersion. In agreement with an
earlier study \citep{2005A&A...443.1067N}, we found that
fully-developed disc turbulence at a level consistent with
observational constraints \citep[see e.g.][and references
therein]{1998ApJ...501L.189A} would cause large-scale diffusion of
planetesimals over distances of several ${\rm au}$ (in contradiction
of solar system constraints) and induce their collisional destruction.

However, as originally contemplated by \citet{1965Icar....4..494P},
sustaining magnetic fields within the protosolar nebula requires the
gas to be sufficiently ionised. This was later pointed out in the
context of magnetised disc turbulence by \citet{1972IPST..........S}.
In the wake of its discovery in the context of accretion discs, the
linear development of the MRI was studied in detail for the case of
ion-neutral drift \citep{1994ApJ...421..163B}, and Ohmic diffusion
\citep{1996ApJ...457..798J,1999ApJ...515..776S}.  The requirements for
non-linear turbulence to be sustained have been studied by means of
direct simulations for the case of Ohmic resistivity
\citep{2000ApJ...530..464F}, and ambipolar diffusion
\citep{1998ApJ...501..758H,2011ApJ...736..144B}. Because of the
stabilising effect of epicyclic oscillations, a purely hydrodynamic
flow is expected to be stable via the Rayleigh criterion
\citep{1917RSPSA..93..148R}. Lacking a plausible source for a
sufficient number of free electrons, the flow is likely to remain in a
laminar state.

Protoplanetary discs in the T~Tauri stage are cold and dense, leading
to the above mentioned low levels of ionisation
\citep{1983PThPh..69..480U}.  Based on the fact that the dominant
ionisation sources are likely external to the disc itself,
\citet{1996ApJ...457..355G} proposed a layered model for partially
ionised accretion discs: While the disc surface layers are well
ionised by stellar X-rays, and sustain a turbulent accretion flow
caused by the MRI, the shielded disc interior is expected to harbour a
`dead zone' where the flow remains laminar implying negligible levels
of accretion. Such layered discs were first investigated numerically
by \citet{2003ApJ...585..908F}, who included a magnetic diffusivity
varying with height. Their box simulations demonstrated that the flow
within the dead zone-region maintains a modest Reynolds stress due to
waves being excited by the active layers.

Owing to the high gas density in the bulk of the disc and the embedded
sub-micron dust grains, a number of physical effects may significantly
affect the overall ionisation balance. This holds true for both PPDs
in general, and the protosolar nebula in particular
\citep{1981PThPS..70...35H}. The effect of dust on charge carriers was
first investigated by \citet{2000ApJ...543..486S}. In a series of
papers, \citeauthor{2006A&A...445..205I} studied the effect of small
dust grains and different chemical reaction networks
\citep{2006A&A...445..205I}, and the role of turbulent mixing
\citep{2006A&A...445..223I}. An additional complication arises when
one considers the Hall term in Ohm's law \citep{1999MNRAS.307..849W},
introducing a dichotomy with respect to the mutual orientation of any
vertical magnetic flux and the rotation axis
\citep{2011arXiv1103.3562W}. A study of the effect of the Hall term on
the MRI turbulence saturation amplitude by \citet{2002ApJ...577..534S}
found it to be little changed for the range of Hall parameters they
investigated.

Studies of the effects of ohmic dissipation determined by
time-dependent reaction networks coupled to the MHD evolution have
been presented by \citet{2007ApJ...659..729T} and
\citet{2008A&A...483..815I}. These works have shown that enlivening
the dead zone via turbulent mixing of charge carriers is only
efficient in the absence of small dust grains. However, if a
significant population of micron sized grains is present, the
adsorption time scale of free electrons is short enough to maintain
the dead zone \citep{2008ApJ...679L.131T,2011arXiv1107.2935B}. 
\citet{2005ApJ...628L.155I} have suggested that fast moving electrons
generated in magnetic reconnection events may provide a source of
ionisation that can remove or modify the dead zone, but detailed
calculations demonstrating the feasibility of this idea have yet to be
done.

Building on this line of work, we have extended our study from Paper~I
to the case of stratified PPDs. This includes a fiducial model being
fully MRI active, and two models including a magnetically inactive
midplane region \citep*[hereafter ``Paper~II'']{2011MNRAS.415.3291G}.
The key results of this study were that the stirring of the
planetesimals' velocity dispersion was much reduced in the dead zone,
as was their radial diffusion -- satisfying the observational
constraint that radial mixing of asteroids has probably not occurred
over distances $\simgt 0.5\au$ \citep{1982Sci...216.1405G}.

This possibly allows for the continued growth of planetesimals rather
than destruction or erosion during collisions, and it was suggested
that dead zones may provide safe havens for planetesimals.  One
perceived limitation of Paper~II was that we restricted ourselves to
the case of a minimum mass protosolar nebula. While one would expect
the strength of the stirring to scale approximately linearly with the
disc mass (all other things being equal), the relative level of
density fluctuations might be affected by the extended width of the
dead zone. It is the purpose of this paper, to elucidate on the net
effect of these two counteracting dependencies, and moreover to
examine the influence of varying the imposed magnetic field strength.

This paper is organised as follows: the physical model and numerical
methods are briefly recapitulated in Section~\ref{sec:methods}; for a
more thorough description, we refer the reader to Paper~II. Simulation
results are presented in three parts: in Section~\ref{sec:disc_mass}
we study the dependence on the disc mass/column density, and in
Section~\ref{sec:net_flux}, we show results on the variation with the
external magnetic field permeating the disc vertically. A modified
scaling relation to encompass all the obtained results is then
developed in Section~\ref{sec:drho_trq}. Implications for the scope
and applicability of our results for planet formation theory are
discussed in Section~\ref{sec:discussion}, and conclusions are drawn
in Section~\ref{sec:conclusions}.


\section{Methods}
\label{sec:methods}

The simulations presented in this paper are based on our fiducial dead
zone model D1 discussed in Paper~II. We again solve the standard
equations of resistive MHD in a locally corotating, Cartesian
coordinate frame ($\xx,\yy,\zz$), adopting the shearing box formalism
\citep[see][and eqns. in sect.~2 of
Paper~II]{2007CoPhC.176..652G}. Embedded in the dynamically evolving
plasma are swarms of 25 non-interacting massive test particles, which
are affected by disc gravity and the inertial forces in the local
frame of reference. All simulations utilise the \NIII code
\citep{2004JCoPh.196..393Z}; further modifications to the solver are
documented in sect.~2.4 of Paper~II.

The base state of our model is given by isothermal stratification in a
fixed gravitational potential $\Phi(z)$. Unlike in earlier studies
\citep{2003ApJ...585..908F,2007ApJ...670..805O}, and to allow for
sufficiently wide MRI-active regions, model D1 used a box size of
$5.5$ scale heights, $H$, on each side of the disc
midplane.\footnote{With our definition of $H$, the equilibrium density
profile is proportional to $\exp(-z^2/(2 H^2))$, i.e. $H$ is identical
to the `$h$' in \citet{2011ApJ...742...65O}.} Based on our previous
analysis of the influence of box size on gravitational stirring
(Paper~I), we furthermore maintain a horizontal box size of
$3\times12\,H$, permitting the excitation of spiral density (SD) waves
\citep{2009MNRAS.397...52H,2009MNRAS.397...64H,2011arXiv1109.2907H}.

\begin{table}\begin{center}
\caption{Overview of simulation parameters, including model D1 from
  Paper~II. The fiducial model D1.1 is identical to D1 and serves as a
  control for the changes described in Sect.~\ref{sec:ind}
  below.\label{tab:models}}
\begin{tabular}{lcccccc}\hline
& $\rho_{\rm mid}$ & $\Sigma\ [\frac{\g}{\cm^2}]$ 
& domain $\ [\,H\,]$& resolution \\[4pt]
\hline
D1 / D1.1 & 1 & 134.6 & $3\tms12,\,\pm 5.500$& $\ 72\tms144\tms264$ \\
D1.2      & 2 & 269.2 & $3\tms12,\,\pm 5.667$& $\ 72\tms144\tms272$ \\
D1.4      & 4 & 538.4 & $3\tms12,\,\pm 5.833$& $\ 72\tms144\tms280$ \\
\hline
\end{tabular}
\end{center}
\end{table}

Because the resistivity $\eta$ depends on a physically motivated
ionisation model involving chemical rate equations, we have to
introduce a unit system to convert code units into standard units. To
maintain continuity with Paper~II, we adopt the same basic disc model,
which is similar to the widely used ``minimum-mass'' protosolar nebula
\citep{1981PThPS..70...35H}. At $5\au$ the Hayashi model has a column
density $\Sigma = 150\g\cm^{-2}$ and a disc opening angle
$H/r=0.047$. We adopt similar values $\Sigma=135\g\cm^{-2}$ and
$H/r=0.05$, giving a temperature $T=108\K$, and an isothermal sound
speed of $c_{\rm s} =667\ms$ (also cf. Tab.~\ref{tab:params}).
Because the minimum-mass assumption does in fact only specify a lower
limit to the expected mass of the protosolar nebula at the age when
planetesimal formation likely occurred, we here study further models
with double and four-times the midplane density $\rho_{\rm mid}$ (see
Tab.~\ref{tab:models}). The ionisation fluxes for X-rays (XR),
short-lived radionuclides (SR), and cosmic rays (CR) are unchanged
from model D1, i.e., XR and CR fluxes are the nominal values, and SRs
are enhanced by a factor of ten; see Sect.~\ref{sec:ionisation} for
further details. For studying the influence of the net vertical field
(NVF), we performed additional runs D1-WF (with a weak field), D1-NVF
a/b (with varying field), and two extra models based on D1.4 with
weaker fields. We adopt a standard resolution of 24 grid cells per $H$
in the radial, and vertical directions; for the azimuthal direction we
use a reduced grid spacing of $12/H$. A basic convergence check is
provided in Appendix~\ref{sec:resol}.

\begin{figure}
  \center\includegraphics[width=0.8\columnwidth]{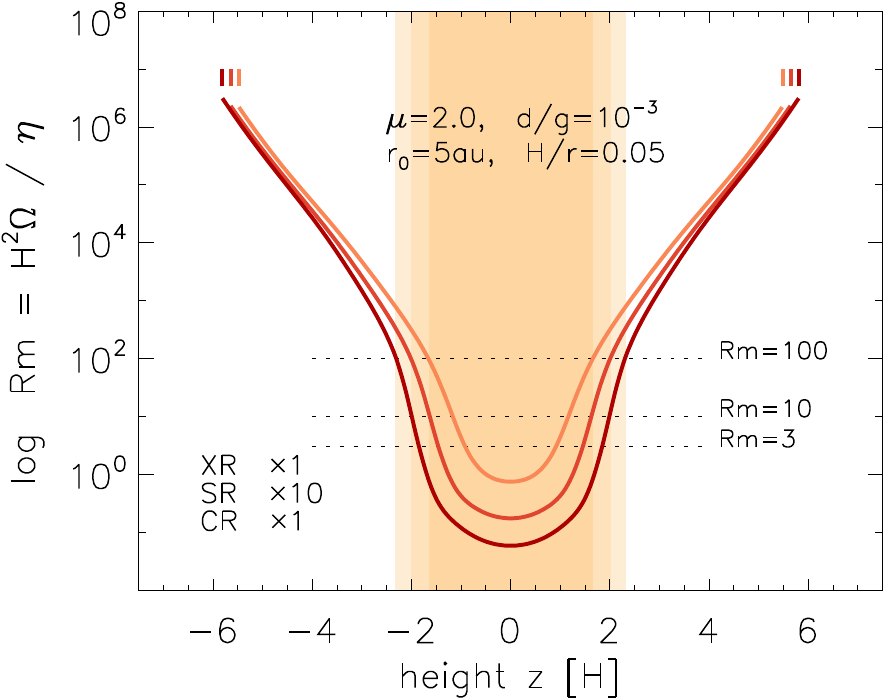}
  \caption{Vertical $\eta$~profiles for the runs presented in this
    paper.  We plot the magnetic Reynolds number,
    $\Rm=\Omega\,H^2\,\eta(z)^{-1}$ determined from our ionisation
    model. The three line represent the initial models for runs D1,
    D1.2, and D1.4, respectively (cf. Tab.~\ref{tab:models}); short
    vertical bars illustrate the varying domain size $L_z$.}
  \label{fig:ppd_strat}
\end{figure}

\begin{table}\begin{center}
\caption{List of model parameters. Note also that magnetic field
  values for each run are listed in the third column of
  table~\ref{tab:sim_results}.\label{tab:params}}
\begin{tabular}{lccc}\hline
parameter               & symbol         & value      & unit\\\hline
orbital location        & $a$            & 5.0        & $\rm au$ \\
disc aspect ratio       & $h$            & 0.05       & \\
local temperature       & $T$            & 108        & $\rm K$ \\
mean molecular weight   & $\bar{\mu}$    & 2.0        & $\rm amu$ \\
gas/dust ratio          &                & 0.001      & \\
dust particle size      &                & 0.1        & $\mu{\rm m}$ \\
dust grain density      & $\rho_{\rm d}$ & 3.0        & $\rm g\,cm^{-3}$ \\
disc column density     & $\Sigma$       & 135 -- 538 & $\rm g\,cm^{-2}$ \\
magnetic field strength & $\bar{B}_z$    & 2.68 -- 43 & $\rm mG$ \\
\hline
\end{tabular}
\end{center}
\end{table}

\subsection{Initial and boundary conditions} 

A primary aim of this work is to construct discs of varying mass in
which the active zone remains essentially unchanged in its mass and
level of turbulence in each model, with the only change being in the
mass and vertical thickness of the dead zone. This component of our
study then becomes one in which we examine the effect of varying the
mass of the dead zone keeping all other parameters fixed (see
Tab.~\ref{tab:params} for reference). The numerical set-up we adopt
for the models of varying disc mass is designed to achieve this aim.

The initial magnetic field is a superposition of two components, a
standard zero-net-flux (ZNF) configuration $B_z(x)\sim
\sin(2\pi\,x/L_x)$, and a weak additional vertical net-flux (NVF)
component representing an external or dynamo-generated field. We vary
the strength of the initial ZNF flux to produce a plasma parameter
(i.e. the ratio of thermal to magnetic pressure) $\beta_{\rm
p}\simeq50$ independent of the midplane density (and hence pressure)
of the three studied models. In contrast, we keep the absolute
strength of the external net vertical flux (which is conserved during
the simulations) fixed at a value corresponding to $B_0=10.73\mG$ in
physical units, resulting in a weaker relative field in the midplane
of models with higher mass loading. Note that the NVF component leads
to a plasma parameter varying with height.  As a consequence, the
region of intermediate field strength, i.e. the vertical range, where
the most unstable nonaxisymmetric MRI modes fit into the box, moves up
in the disc. We expect the resulting active regions to be of similar
vertical extent irrespective of the mass loading, but separated by a
wider dead zone in the case of heavier discs as is illustrated by
shaded areas in Fig.~\ref{fig:ppd_strat}. To maintain near-identical
active zones across all runs, we adjust the vertical box size
accordingly (cf. Tab.~\ref{tab:models}).

The simulations that examine the influence of varying the strength of
the net vertical magnetic field strength (models D1-NVFa and D1-NVFb)
are initiated with specific values of the magnetic field
($B_0=2.68\mG$ and $B_0=10.73\mG$, respectively), and after every 40
orbits the field strength is incremented by factors of two up to some
maximum value.  Although we only adopt two run labels for these
simulations, these simulations in fact explore the influence of six
different external magnetic field strengths ranging between $2.68\mG$
and $86\mG$.

As discussed in more detail in Paper~II, for the fluid variables, we
use vertical boundary conditions allowing material to leave the box
but prevent inflow. Mass loss associated with the outflow boundaries
is compensated by continually re-instating the initial density profile
in each grid cell by means of an artificial mass source term
\citep[cf.][]{2009A&A...498..335H}. Because of the low density in the
halo, the relative change in the disc mass due to outflow is very low
(on the order of $10^{-4}$ per orbit). We have verified that the
additional term does not affect the obtained stirring amplitudes.  As
in our previous work, we compute a locally and temporally varying
magnetic diffusivity $\eta({\bf x},t)$. This adds to the level of
realism achieved in earlier simulations of dead zones by
\citet{2003ApJ...585..908F} and \citet{2007ApJ...670..805O}, who used
static diffusivity profiles. Here we briefly re-capitulate the
physical ionisation model we implement and refer the reader to
section~2.3 of Paper~II for a more detailed discussion.

\subsection{The diffusivity model} 

Because of the expected dominance of small dust grains
\citep{2000ApJ...543..486S,2006A&A...445..205I}, we decide to avoid
following the detailed non-equilibrium chemistry and adopt a
simplified approach for the gas-phase reactions. Assuming that
recombination happens much faster than any dynamical mixing timescale
in our system, we update $\eta$ according to a precomputed table
derived from the reaction network in \texttt{model4} of
\citet{2006A&A...445..205I}. When computing the resistivity, we
include the contributions of all the charged species following
\citet{2007Ap&SS.311...35W}, eqns.~(21)-(31). For the free parameters,
we assume the same values described in Paper~II, i.e. dust grains of
size $0.1\mu{\rm m}$ and with density $3\g\perccm$ and a dust-to-gas
mass ratio of $10^{-3}$ (such that we tacitly assume that 90 percent
of the solid grains have already grown to become larger bodies). The
gas-phase abundance of Magnesium is taken to be depleted by a factor
$10^{-4}$ compared to its solar value (with the remainder assumed to
be bound-up in grains). The key factor governing the diffusivity
$\eta$, is the local ionisation rate $\zeta({\bf x},t)$, which is
computed from the external irradiation by evaluating column densities
to both the upper and lower disc surfaces. Because of the expected
radial density features, related to the excited spiral waves, this is
done on a per-grid-cell basis.

\subsubsection{Ionisation sources} 
\label{sec:ionisation}

The ionisation model is founded on the work of
\citet{2009ApJ...703.2152T}, who studied how stellar X-rays,
radionuclides, and energetic protons can influence the shape and
extent of a possible dead zone via their effect on Ohmic diffusion.
Implementing the model of \citet{2009ApJ...703.2152T}, we focus on
stellar X-rays (XR), and interstellar cosmic rays (CRs) as the prime
sources of ionisation. As a representative value, we chose $L_{\rm XR}
\simeq 2 \times 10^{30}\,{\rm erg\,s}^{-1}$ as suggested by
\citet{2000AJ....120.1426G}. Applying a simple fit to the Monte-Carlo
radiative transfer calculations of \citet{1999ApJ...518..848I}, we
approximate the ionisation rate due to X-rays by
\begin{equation}
  \zeta_{\rm XR} = 2.6\tms10^{-15}\s^{-1}\ 
  \left[ {\rm e}^{-\Sigma_a/\Sigma_{\rm XR}}
       + {\rm e}^{-\Sigma_b/\Sigma_{\rm XR}}
  \right]\,r_{\au}^{-2}\,,
\end{equation}
with $\Sigma_a$ and $\Sigma_b$ the gas column densities above and
below a given point, and $\Sigma_{\rm XR}=8.0\g\cm^{-2}$ the assumed
X-ray absorption depth. For the vertical attenuation of interstellar
CRs illuminating the disc surfaces, we adopt the formula given in
\citet{2009ApJ...690...69U}:
\begin{equation}
\zeta_{\rm CR} = 5\tms10^{-18}\s^{-1}
  \ {\rm e}^{-\Sigma_a/\Sigma_{\rm CR}}\ 
  \left[ 1+\left(\frac{\Sigma_a}{\Sigma_{\rm CR}}\right)^\frac{3}{4}
  \right]^{-\frac{4}{3}}
  +\ \dots\,,
\end{equation}
where $\Sigma_{\rm CR}=96\g\cm^{-2}$ is the cosmic ray attenuation
depth \citep{1981PASJ...33..617U}, and dots indicate the corresponding
contribution from the second column density $\Sigma_b$. We further
include an ambient ionisation due to the decay of short-lived
radionuclides (SR). As already done for model D1, and to somewhat
limit the dynamic range in $\eta$, the related ionisation is chosen
$10\times$ the nominal value of $\zeta_{\rm SR}=
3.7\tms10^{-19}\s^{-1}$ quoted in \citet{2009ApJ...703.2152T}.

\subsection{The induction equation} 
\label{sec:ind}

When simulating protoplanetary discs with dead zones using the
shearing box approximation, there are two issues that arise which must
be addressed to increase the feasibility and realism of the
models. The first occurs because we are solving an advection-diffusion
equation in time-explicit fashion. The time step size is restricted by
the time for diffusion across one grid cell, and because this quantity
scales with the square of the grid spacing, the presence of large
diffusion coefficients can render the computational effort prohibitive
in highly resolved runs. This was addressed in our previous paper by
limiting the dynamic range in $\eta$ and applying the technique of
temporal sub-cycling to the diffusive part of the induction equation
(cf. sect.~2.3.2 in Paper~II).  With the more extended dead zones of
models D1.2 and D1.4, the time scales become even more disparate, and
we have developed a modified scheme to deal with this, described in
detail in Appendix~\ref{sec:ind_mod}.

The second issue relates to the generation of strong azimuthal
magnetic fields through the winding up of net radial fields in the
box. Although our simulations begin with zero net radial magnetic
field in the computational domain, advection of horizontal field
components through the open vertical boundary leads to the generation
of net radial fields that may diffuse into the dead zone. Once there,
Keplerian shear may wind them up to generate strong azimuthal fields
that subsequently leak back into the active regions, modifying the
turbulence there.  This issue has been discussed by
\citet{2008ApJ...679L.131T}, where it was noted that the winding of
fields generates a strong radial gradient in the azimuthal field
strength due to the Keplerian rotation profile in global discs,
leading to radial diffusion of the strong fields that limits their
growth. This radial diffusion cannot occur in a shearing box due to
the uniform shear and periodic boundary conditions (but has been
observed in the global simulations presented by
\citet{2010A&A...515A..70D}), so we have developed an approximate
scheme designed to crudely mimic the radial diffusion of fields in
global discs. We refer to this scheme as super-box scale diffusion,
and we describe its implementation in Appendix~\ref{sec:ind_mod}.


\section{Dependence on disc mass}
\label{sec:disc_mass}

In Paper~II, we restricted our analysis to the fiducial case of a
minimum mass protosolar nebula, and this naturally begs the question
how the results will be affected if one looks at more massive discs.
A major aim of this study is to compute disc models with different
masses/surface densities that have very similar active layers in terms
of mass and turbulent activity. The main difference between the models
will then be the physical depth and mass of the dead zone. Density
waves that are excited in the active layer will propagate into the
dead zone and induce gravitational stirring of the planetesimals there
through the density fluctuations that they generate. A key question is
how this stirring depends on disc mass.

\begin{figure*}
  \includegraphics[height=0.88\columnwidth]{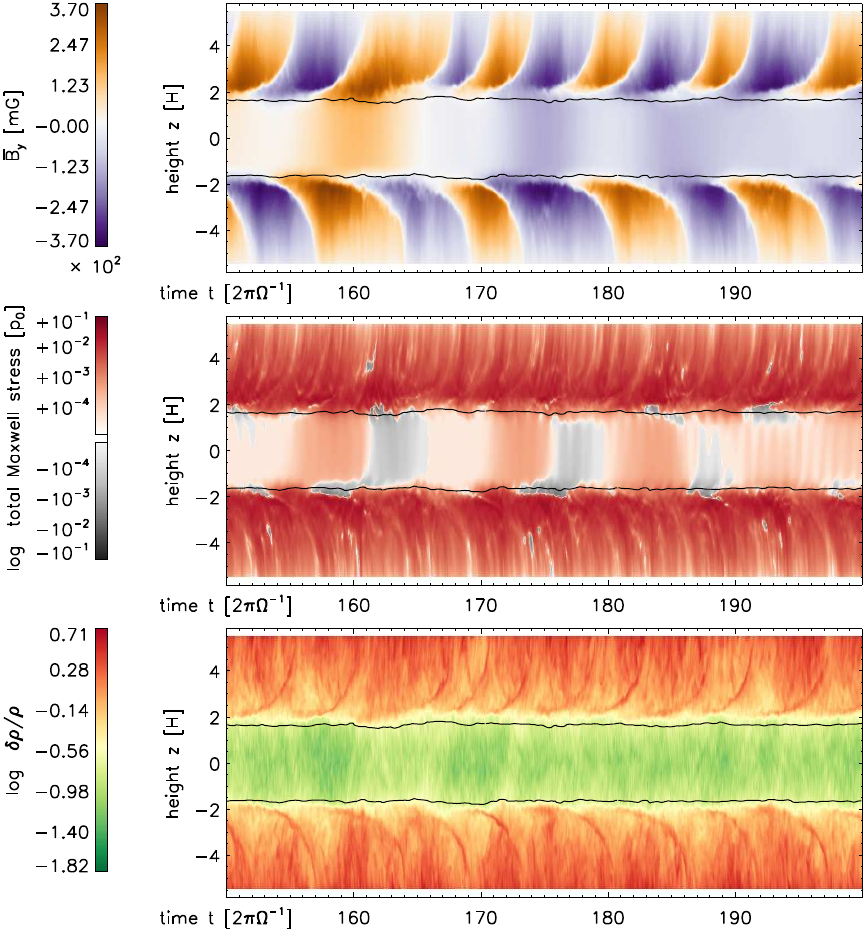}\hspace{0.5ex}%
  \includegraphics[height=0.88\columnwidth]{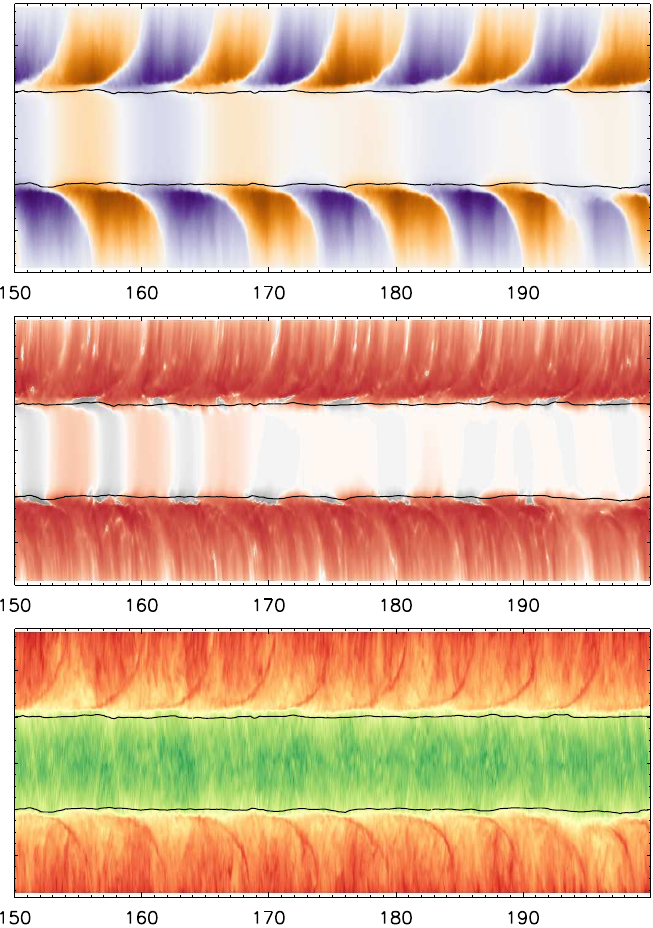}\hspace{0.5ex}%
  \includegraphics[height=0.88\columnwidth]{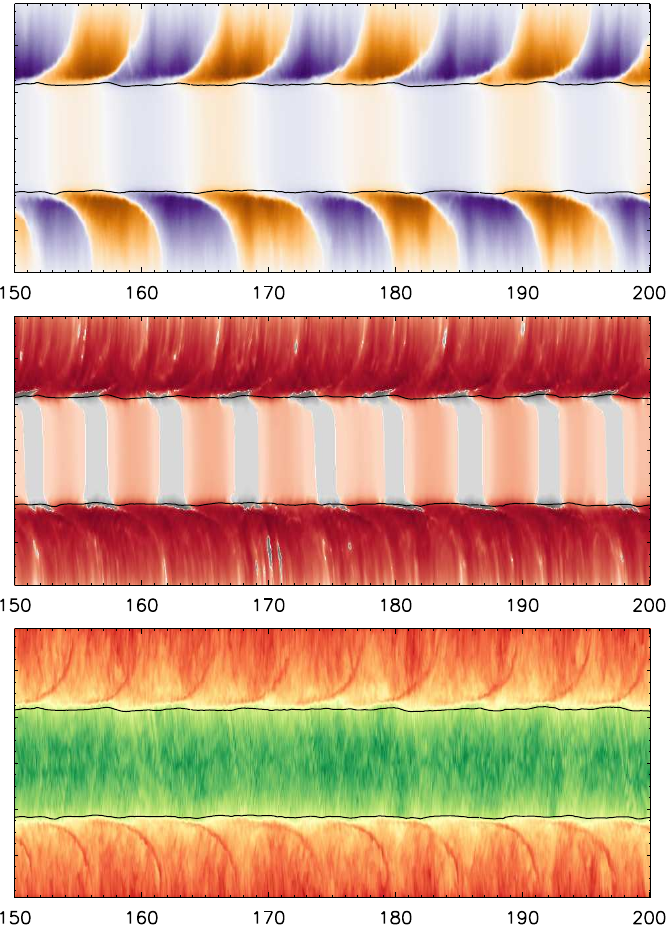}
  \caption{Space-time evolution of the mean toroidal field
    $\bar{B}_\phi$ (upper row), total Maxwell stress
    $M_{r\phi}=-\rms{B_R\,B_\phi}$ (middle row), and relative density
    fluctuation $\delta \rho/\rho$, (bottom row). The different panels
    show runs D1.1 (left), D1.2 (centre), and D1.4 (right). The
    interface between the live and dead zones is indicated by a solid
    line representing the criterion $\El\equiv v_{\rm
      A}^2/(\Omega\eta)=1$. Note the trends of a wider dead zone, and
    weaker \emph{relative} density perturbations when going to higher
    densities, while most disc properties remain virtually unchanged.}
  \label{fig:spctm}
\end{figure*}

\subsection{Scaling relations for linear density waves} 
\label{sec:linear_waves}

Based on the results of simulations presented in paper II, we expect
the disc models described below to have the following
midplane-symmetric three-layer structure: a non-turbulent dead zone
located at and above the midplane in which the MRI is quenched by
Ohmic resistivity, on top of which lies a turbulent region in which
the MRI operates unaffected by resistivity. At disc altitudes above the
MRI-active region lies a magnetically dominated halo in which the
field strength is too large for the MRI to operate because unstable
modes do not fit within the available vertical extent. Density waves
excited at the interface between the dead zone and MRI-active layer
will propagate into the dead zone, creating density fluctuations at
the midplane.

We denote the density at the active/dead zone interface as $\rho_{\rm
int}$.  The disc models of different mass we compute are designed to
generate discs in which very similar active layers lie above and below
dead zones of different mass, and for which the midplane density,
$\rho_{\rm mid}$, scales with the disc mass. We expect each of our
models therefore to have very similar values of the density,
$\rho_{\rm int}$, and perturbed velocity, $\delta v_{\rm int}$, at the
dead/active zone interface. Linear density waves that travel sonically
into the dead zone should conserve energy as they propagate, such that
we expect $\rho_{\rm mid} (\delta v_{\rm mid})^2 \simeq \rho_{\rm int}
(\delta v_{\rm int})^2$, where $\delta v_{\rm mid}$ is the perturbed
velocity at the midplane. Relative density fluctuations for isothermal
sound waves satisfy the relation
\begin{equation}
 \frac{\delta\rho}{\rho} = \frac{\delta v}{c_{\rm s}}
 \label{eq:delta_rho}
\end{equation}
such that relative density fluctuations at the disc midplane
should obey the scaling
\begin{equation}
  \left(\frac{\delta \rho}{\rho}\right)_{\rm mid}
  \propto \frac{1}{\sqrt{\rho_{\rm mid}}}.
\label{eq:delta_rho_mid}
\end{equation}
Translated into absolute density fluctuations, this implies a scaling
$\delta \rho_{\rm mid} \propto \sqrt{\rho_{\rm mid}}$. It is
reasonable to expect the fluctuating torque, $\Gamma_y$, induced by
density fluctuations near the midplane will scale linearly with
$\delta \rho_{\rm mid}$, leading to the naive expectation that
stochastic torques will scale according to $\Gamma_y \propto
\sqrt{\rho_{\rm mid}}$.

At the time of writing an extensive study of MRI turbulence in layered
accretion discs appeared in print \citep{2011ApJ...742...65O}.  For a
broader discussion about scaling relations in disc models with dead
zones we refer the reader to this recent publication.

\subsection{Disc evolution} 
\label{sec:disc_evol}

As in sect.~3.2 of Paper~II, we will start our discussion by looking
at the evolution of the gas disc as shown in
Figure~\ref{fig:spctm}. Owing to the additional magnetic diffusion
term (cf. Sect.~\ref{sec:super_box}), the overall evolution of model
D1.1 (left hand panels) appears somewhat less intermittent than the
identical setup from model D1 in Paper~II. This is also true for the
heavier models D1.2 (centre) and D1.4 (right hand panels).

The different models in fact appear very similar, e.g., in terms of
the cycle frequency of the field reversals and the overall field
amplitude. Also the stresses and relative density fluctuations within
the MRI-active regions are nearly identical. The only apparent trend
is seen in the relative density fluctuations in the midplane region,
which fall off with mass loading. The relative amplitudes, as listed
in column 7 of Tab.~\ref{tab:sim_results}, are 0.091, 0.058, and
0.038, respectively, which implies a scaling with the midplane density
to the power of $\sim -0.63\pm0.01$; this is in decent agreement with
the expected scaling $(\delta \rho_{\rm mid}/\rho_{\rm mid}) \propto
\rho_{\rm mid}^{-1/2}$ discussed above in
Sect.~\ref{sec:linear_waves}. The discrepancy can partly be ascribed
to the varying $\delta\rho\,|_{\El=1}$ at the layer interface (cf.
Sect.~\ref{sec:disc_struct} below), which shows a slight trend -- with
a power $\sim -0.05\pm0.01$ -- towards weaker absolute fluctuations in
heavier discs. Overall, the results indicate that the propagation of
density waves from the transition region between the active and dead
zones occurs without significant dissipation of energy.

Because our model assumes that the main source of ionisation is
exterior to the disc surface, the width of the dead zone depends on
the column density of the shielding material. The extent of the dead
zone as a function of various ionisation sources was studied in detail
by \citet{2009ApJ...703.2152T}. In our simulations, we determine the
transition region via the criterion that the Elsasser\footnote{We
note, in passing, that this definition is formally identical to what
has been called a Lundquist number (or even magnetic Reynolds number)
in related publications. Irrespective of its name, $\El<1$ serves as a
robust indicator for suppression of the fastest-growing MRI mode
\citep[see][]{2010ApJ...716.1012P}.}  number $\El\equiv v_{\rm
A}^2/(\Omega\eta)$ equals unity (indicated by a solid black line in
Fig.~\ref{fig:spctm}).  This yields a time-averaged half thickness of
1.66$\pm$0.04, 2.02$\pm$0.03, and 2.33$\pm$0.03$\,H$ for models
D1.1-4, respectively. In other words, doubling the disc mass increases
the extent of the dead zone by a third of a pressure scale height
$H$. As can be seen from Fig.~\ref{fig:spctm}, the extent of the
different layers is remarkably stationary. As a result, the dead zone
thickness only fluctuates on the per-cent level. This has, of course,
to be taken with a grain of salt. External conditions (e.g., gas
inflow from the parent molecular cloud, stellar X-ray luminosity,
intensity of arriving CRs, $\dots$) in a realistic protoplanetary disc
are likely to vary strongly both in time and space. However, the
relative quiescence of our experimental setup will allow us to
quantify the secular evolution of the embedded planetesimals more
accurately.

\subsection{Disc vertical structure} 
\label{sec:disc_struct}

\begin{figure}
  \center\includegraphics[width=0.9\columnwidth]{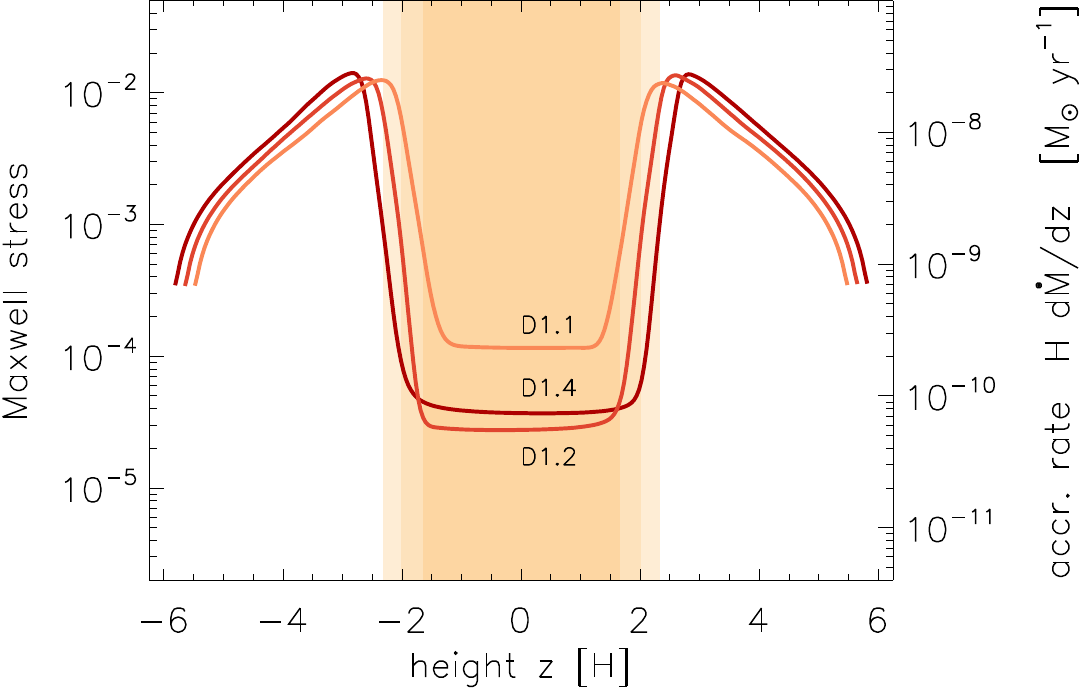}
  \caption{Time-integrated Maxwell stresses for models D1.1-4 as
    labelled. The vertical displacement of the curves demonstrates
    nicely why all models produce the same mass accretion rate. Shaded
    areas indicate the extent of the dead zone in each
    model.\label{fig:Maxw_z}}
  \center\includegraphics[width=0.9\columnwidth]{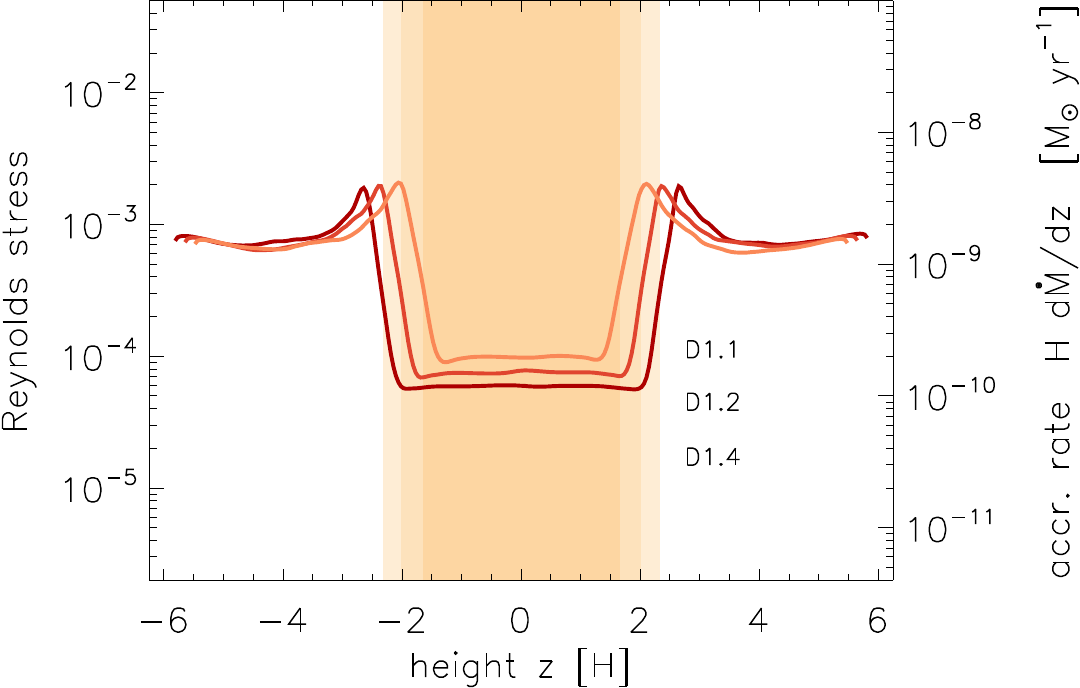}
  \caption{Time-integrated Reynolds stresses for models D1.1-4. The
    same vertical displacement as in Fig.~\ref{fig:Maxw_z} is
    seen. The peaks in the stress sits just outside the transition
    region between the active and inactive layers
    (shaded).\label{fig:Reyn_z}}
\end{figure}

Having assured ourselves of the stationary nature of the flow, we can
now proceed and look at the characteristic structure of the layered
disc. In Figure~\ref{fig:Maxw_z}, we plot the time-averaged (20-220
orbits) Maxwell stress $M_{r\phi}=-\rms{B_r\,B_\phi}$ for the three
models D1.1-4, of varying column density. As discussed in the last
section, owing to the random diffusion of flux into the midplane
region, there is no systematic to the residual level within the dead
zone. As noted earlier, we expect the MRI-active zone to move away
from the midplane for the higher density models, and this trend is
clearly seen. As a result, we infer very similar vertically-integrated
mass accretion rates of $7.90\times 10^{-8}$, $7.97\times10^{-8}$, and
$7.86\times10^{-8}\Msun\yr^{-1}$ for runs D1.1-4, respectively. This
is in-line with the volume-averaged values stated in column 4 of
Tab.~\ref{tab:sim_results} and demonstrates that higher mass loading
does not affect the level of activity or mass in the active zone, but
merely the vertical extent and mass of the dead zone, such that the
nonlinear evolution of the disc models arising from the specified
initial conditions is as intended.

Figure~\ref{fig:Reyn_z} shows the same shift away from the midplane
for higher disc mass in the hydrodynamic stresses
$\rms{\rho\,v_r\delta v_\phi}$. The associated, vertically integrated
mass accretion rates are $1.28\times10^{-8}$, $1.33\times10^{-8}$, and
$1.41\times10^{-8}\Msun\yr^{-1}$ for models D1.1-4, respectively,
i.e. the bulk of the transport is due to magnetic stresses. Within the
dead zone, the Reynolds stresses scale with the gas density but this
trend is very weak at their respective peak position, which lies just
outside the layer interface. This region is of particular interest as
it marks the position in the disc where the turbulent fluctuations
created by the MRI-active layers are most energetic in terms of their
mass loading and hence their associated momenta. Tracing the interface
via the $\El=1$ criterion, we obtain very similar \emph{absolute}
density fluctuations at this line. The respective values are
$\delta\rho\,|_{\El=1}=$ 0.045, 0.043, and 0.042 for models D1.1-4

We conclude that, in all three models, the dead zone sees a very
similar boundary separating it from the active zone, with the only
difference being in its physical separation from the particle swarm
located near $z=0$. In the following section, we will determine the
consequences this has on the gravitational forcing exerted onto the
planetesimals, and examine if the stochastic torques follow the
scaling predicted in Sect.~\ref{sec:linear_waves}.

\subsection{Gravitational torques} 
\label{sec:trq}

It was established in the global simulations of
\citet{2004MNRAS.350..849N} and \citet{2005A&A...443.1067N} that
density fluctuations from developed turbulence lead to stochastic
gravitational forces which can have a significant impact on the
dynamical evolution of embedded planetesimals and protoplanets. We
remark that earlier shearing box simulations of MHD turbulence have
shown the development of axisymmetric pressure bumps or `zonal flows'
\citep{2009ApJ...697.1269J}, which may influence the stirring of
planetesimals. Analysis of our simulation D1.1 suggests that the
midplane density supports a long-lived (20-30 orbits), axisymmetric
density asymmetry across the radial width of the box with amplitude
$\sim 1$\%. This may be an indication that weak zonal flows arise even
in the presence of a dead zone. The role of this in stirring
planetesimals is unclear at present and will the subject of a future
study.

In our previous papers, we have quantified the level of stirring for a
range of disc models. We here perform a similar analysis and find
broad agreement with previous runs. In particular, model D1.1 agrees
well with the earlier run D1 from Paper~II -- we attribute the
discrepancy of about ten per-cent in the torques between D1 (0.06) and
D1.1 (0.054) to the lower level of sampling noise achieved in the new
set of simulations, where a more conservative choice for the
gravitational softening parameter was taken. We note that when
statistically\footnote{As different locations ought to be
indiscriminate in a stochastic sense, this is done by binning the full
data with respect to particle position (at sub-grid resolution) and
then removing the median torque for each bin.} correcting the original
results from D1 for a weak systematic dependence on the particle
position on the grid, excellent agreement can be obtained.

\subsubsection{Torque distribution functions} 
\label{sec:trq_his}

\begin{figure}
  \center\includegraphics[width=0.67\columnwidth]{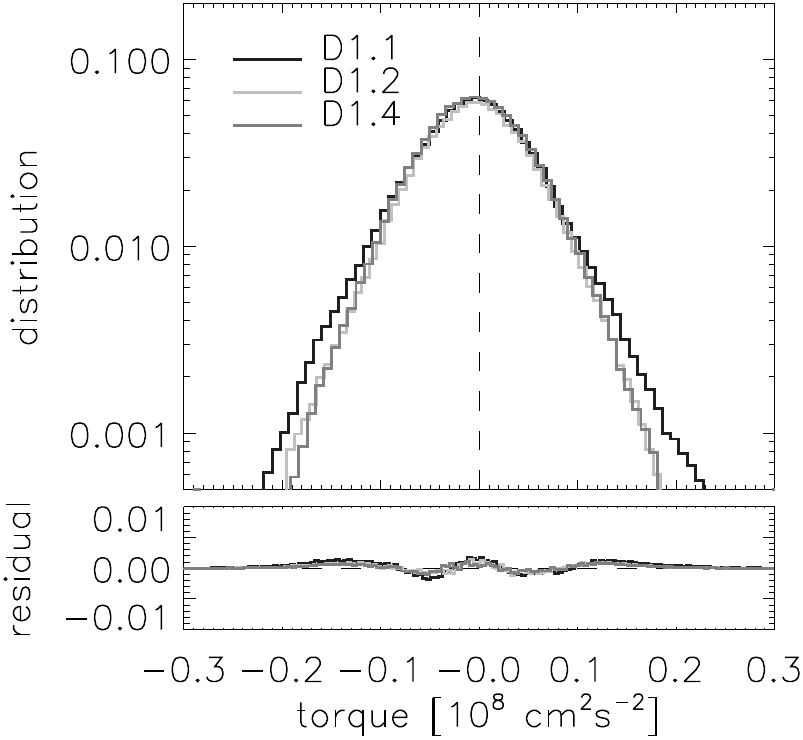}
  \caption{Torque distributions $\Gamma_y$ for the models D1.1-4 with
    varying disc mass. Residuals indicate the deviation from a normal
    distribution.}
  \label{fig:trq_his}
\end{figure}

Torque time series, $\Gamma_{\!y}(t)$, are derived from the
gravitational forces acting on the particles along the $y$~direction
within our simulation box, and are recorded along the evolving
trajectory of the particles. Torque amplitudes,
$\sigma(\Gamma_{\!y})$, for the different runs are obtained from a
Gaussian fit to the histograms as depicted in
Figure~\ref{fig:trq_his}. Averaging over the swarm of 25 particles and
sub-intervals in time allows us to estimate error bars, and we infer
values of $\sigma(\Gamma_{\!y})=$ 0.054$\pm$0.004 for model D1.1, and
0.052$\pm$0.003 for both models D1.2 and D1.4 (these values are also
listed for reference in column eight of
Table~\ref{tab:sim_results}). This demonstrates that the width of the
torque distribution is remarkably insensitive to the column density,
and models D1.1 through D1.4 agree to within five per-cent, which is
somewhat smaller than the error bars. This result is in clear conflict
with the simple scaling law based on conservation of wave energy for
isothermal sounds waves (see Sect.~\ref{sec:linear_waves}).

Naive expectations suggest that fluctuating torques experienced by
solid bodies located near the disc midplane will scale linearly with
absolute density fluctuations there, where we both predict and observe
the scaling $\delta \rho_{\rm mid} \propto \sqrt{\rho_{\rm mid}}$.
Thus, we expect the fluctuating torque amplitude to scale with the
square-root of the disc mass in these simulations. As discussed in
Sect.~\ref{sec:disc_struct}, we find that all models agree on the
relative density fluctuations at the transition between the dead and
active zone, and the absolute and relative density fluctuations at the
midplane are in good agreement with the scaling arguments (also
cf. Fig.~\ref{fig:drho_z}).  This suggests that a more subtle effect
is responsible for modifying the expected linear scaling between the
midplane density fluctuations, $\delta \rho_{\rm mid}$, and the rms
torque fluctuations, $\sigma(\Gamma_{\!y})$, shown in
Figure~\ref{fig:trq_his}. A factor of four change in disc mass between
models D1.1 and D1.4 should lead to a factor of two change in the
torque amplitude, and this would clearly be detectable if present. We
defer a more detailed discussion of this issue until
Sect.~\ref{sec:drho_trq}, where we provide evidence to resolve the
discrepancy between the simulation results and linear scaling.

\subsubsection{Torque autocorrelation} 
\label{sec:trq_acf}

\begin{figure}
  \includegraphics[height=0.45\columnwidth]{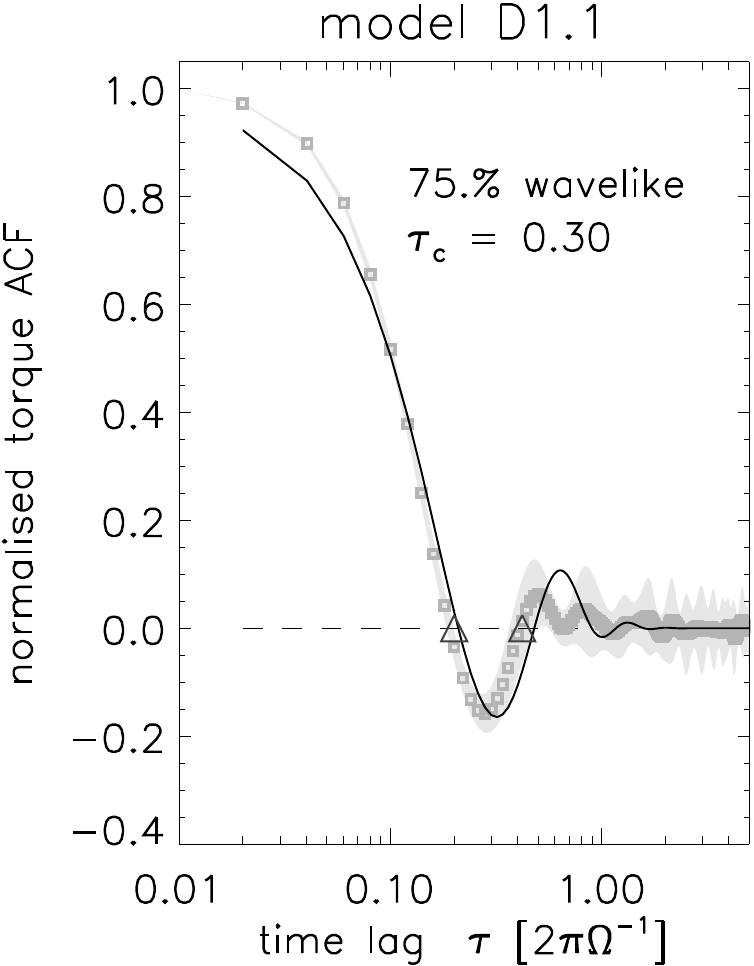}%
  \includegraphics[height=0.45\columnwidth]{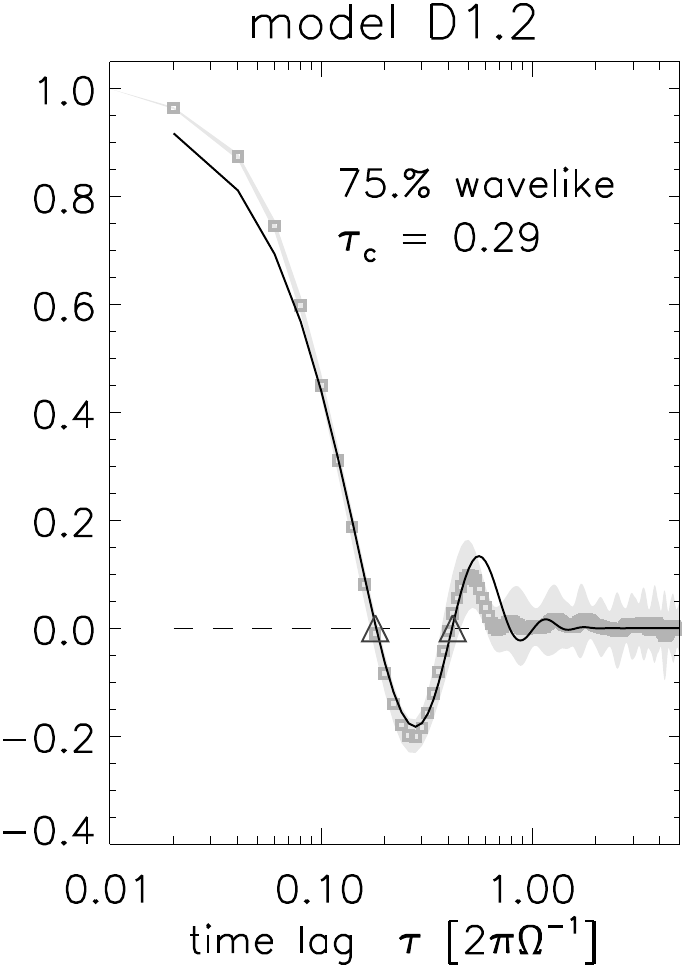}%
  \includegraphics[height=0.45\columnwidth]{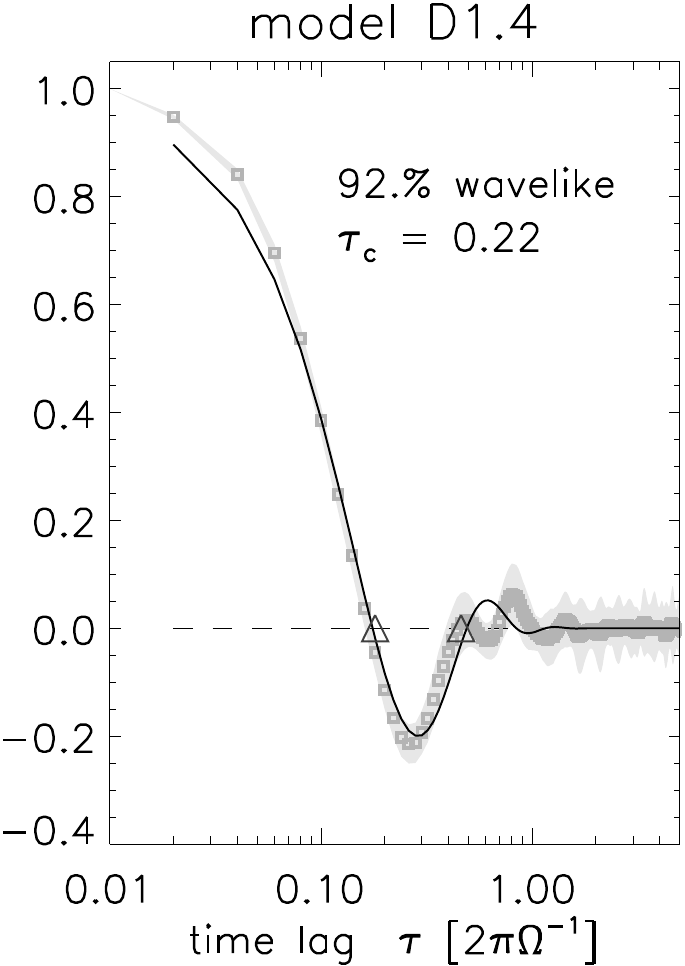}
  \caption{Autocorrelation functions (ACFs) of the torque time series,
    averaged over 25 particles, and 4 sub-intervals in time (shaded
    areas). The first and second zero crossing are indicated by a
    triangle. Labels give the fitted coherence time along with the
    relative amplitude of the wavelike feature.}
  \label{fig:trq_acf}
\end{figure}

How does the modified amplitude of the density waves influence the
temporal coherence of the stochastic torques? We recall from our
previous work that the torque autocorrelation function (ACF) appeared
as a superposition of a truly stochastic part and a modulated
``wavelike'' feature (due to the quasi-periodic passage of density
waves through the particle's location). This dual character, which is
also seen in the corresponding figs.~9 and 15 of
\citet{2007ApJ...670..805O} and \citet{2009ApJ...707.1233Y}, was
described in terms of a formula which we introduced in section 3.3 of
Paper~I:
\begin{equation}
  S_{\Gamma}(\tau) = \left[ (1-a) + a\,\cos(2\pi\,\omega\,\tau) \right]\,
  {\rm e}^{-\tau/\tau_{\rm c}} \label{eq:fit_acf}\,.
\end{equation}
Fitting parameters are the relative amplitude $a$, the frequency of
the wavelike modulation $\omega$, and the coherence time $\tauc$ of
the stochastic envelope. The resulting fits for the models D1.1-4 are
plotted in Fig.~\ref{fig:trq_acf}, where we also state the obtained
values for the relevant fitting parameters. In terms of the parameter
of primary interest, the coherence time $\tauc$, models D1.1 and D1.2
show excellent agreement with model D1 from Paper~II. Note that while
model D1 showed a $\sim55$\% wavelike modulation, this is now
increased to $\sim75$\%. We believe that this relative increase in the
wave feature is, in fact, a reduction of the stochastic component
caused by the lower level of random sampling noise in the new set of
simulations. When correcting for this in model D1, we also obtain a
wavelike amplitude of $\sim75$\%. The modulation is even more
pronounced in model D1.4, which moreover shows a significantly reduced
coherence time of $\tauc=0.22$. One can see by inspection of the
over-plotted model curves that there exists some tension between our
simplistic approach and the actual data. We conclude that a more
accurate determination will require a more thorough understanding of
the ACF shape. Yet, ultimately the derived stochastic description via
$\sigma(\Gamma_{\!y})$ and $\tauc$ can be verified by comparison with
the amount of particle diffusion present in the simulations, as we
shall demonstrate in the next section.

\subsection{Radial diffusion of planetesimals} 
\label{sec:da-grow}

\begin{figure}
  \center\includegraphics[width=\columnwidth]{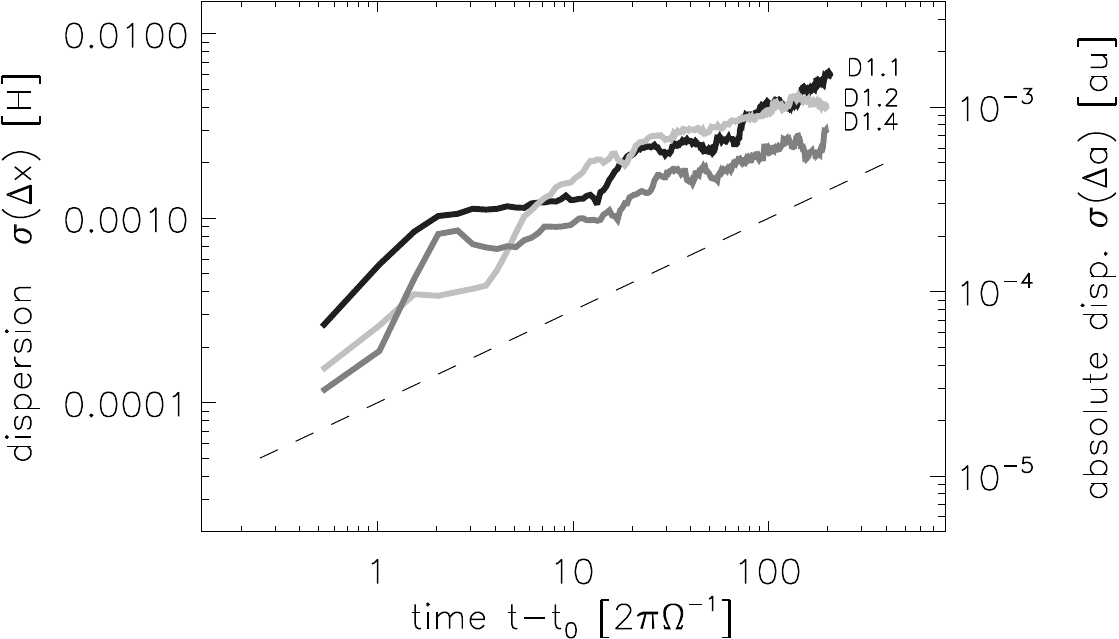}
  \caption{Random-walk behaviour for the rms radial displacement 
    $\sigma(\Delta x)$. The right hand axis shows the absolute dispersion
    $\sigma(\Delta a)$ at $a_0=5\au$. Note that model D1.4 appears to
    show a slightly reduced tendency for particle dispersion.}
  \label{fig:xdisp_cmp}
\end{figure}

Having demonstrated that the gravitational torques acting on embedded
planetesimals can be approximated by a normal distribution, and their
temporal correlations possess a finite coherence time, we can now
exploit these two properties to predict the amount of particle
diffusion based on a Fokker-Planck-type model. This approach was
described in detail in section~6 of Paper~II, and we here briefly
recapitulate the essential ingredients. Particle spreading occurs
because stochastic torques cause diffusion of planetesimal angular
momenta at a rate given by the diffusion coefficient $D_{\rm p} =
\sigma(\Gamma_{\!y})^2\,\tauc$. This leads to an estimated spread in
semi-major axis $a$ after an evolution time, $t$, of
\begin{equation}
  \Delta a = \frac{2 a_0 \sqrt{D_{\rm p}\,t}}{\sqrt{GM\,a_0}},
  \label{eq:delta-a}
\end{equation}
as was derived in eqns. (16)-(19) in Paper~II. This estimate for
$\Delta a$ can then be confronted with a random walk-type fit to the
rms particle displacement in semimajor axis due to diffusion,
$\sigma(\Delta x)$, obtained in the simulations
\begin{equation}
  \frac{\sigma(\Delta x)}{H} = C_{\sigma}(\Delta x) \sqrt{t}\,,
  \label{eq:delta-x-fit}
\end{equation}
corresponding to eqn.~(20) in Paper~II. The rms spread in the local
displacement $\Delta x\equiv \Delta (a - a_0)$ is plotted in
Fig.~\ref{fig:xdisp_cmp} for the different models. Owing to the
relatively small number of particles, the curves display substantial
random fluctuations, yet the general behaviour of a random walk-like
growth emerges, and it can further be seen that the rate of
planetesimal spreading in the runs is similar, as expected from the
earlier discussion concerning the similarity of the stochastic torques
in the three runs. The fitted values obtained for $C_{\sigma}(\Delta
x)$ are $(4.56\pm0.54)\times10^{-4}$ for model D1.1, and
$(4.36\pm0.32)\times10^{-4}$ for model D1.2, in good agreement with
the value for run D1 from Paper~II, which is listed in
Tab.~\ref{tab:sim_results}. As can also be seen from
Fig.~\ref{fig:xdisp_cmp}, the value of $(2.98\pm0.77)\times10^{-4}$
for model D1.4 is somewhat reduced. This is consistent, however, with
the lower inferred coherence time in this case.

To be more specific, let us compare the prediction based
on~(\ref{eq:delta-a}) with the measurement of $C_{\sigma}(\Delta x)$,
which can be related via~(\ref{eq:delta-x-fit}). For model D1.1, we
infer a spread of $\Delta a=0.00161\au$ at a distance of $a_0=5\au$,
and after a run-time of $t-t_0=$200 orbits. This value may readily be
checked by reading it off the right hand axis of
Figure~\ref{fig:xdisp_cmp}. By means of specifying $\tauc$ and
$\sigma(\Gamma_{\!y})$, the diffusion equation predicts $\Delta
a=0.00148\au$, which is only about ten percent smaller than the actual
value. A similar comparison for model D1.2 yields $\Delta
a=0.00154\au$ from Fig.~\ref{fig:xdisp_cmp}, opposed to an estimated
$0.00140\au$ from the diffusion approach, i.e., agreement again to
within ten percent. Finally, for model D1.4 we predict $\Delta
a=0.00122\au$ via (\ref{eq:delta-a}), and observe a value of
$0.00105\au$, which is about 15\% smaller than the estimated
value. Compared to the results from Paper~II, this means a
substantially improved accuracy of the predictions. We attribute this
to the more stationary behaviour of the new simulations effected by
the inclusion of super-box scale dissipative effects
(cf. Sect.~\ref{sec:super_box}). We conclude that the simplified
description of particle spreading in terms of a diffusion equation can
be used as a reliable tool in predicting the level of dispersion in
the immersed planetesimal population within a dead zone.  Moreover,
the good agreement between the estimated and measured values lends
support to the accuracy of the coherence times fitted via
(\ref{eq:fit_acf}).

\subsection{Eccentricity stirring} 
\label{sec:e-growth}

To complete our discussion on how the embedded planetesimals are
affected by the turbulence, we now briefly consider the excitation of
their eccentricity. This has been done in some detail in section~5 of
Paper~II, including a discussion of the long-term evolution and
related saturation mechanisms. In this section, we consider the
results of the simulations directly, and leave discussion about their
implications for planetesimal accretion and planet formation theory
until later in Sect.~\ref{sec:discussion} of this paper.

\begin{figure}
  \center\includegraphics[width=\columnwidth]{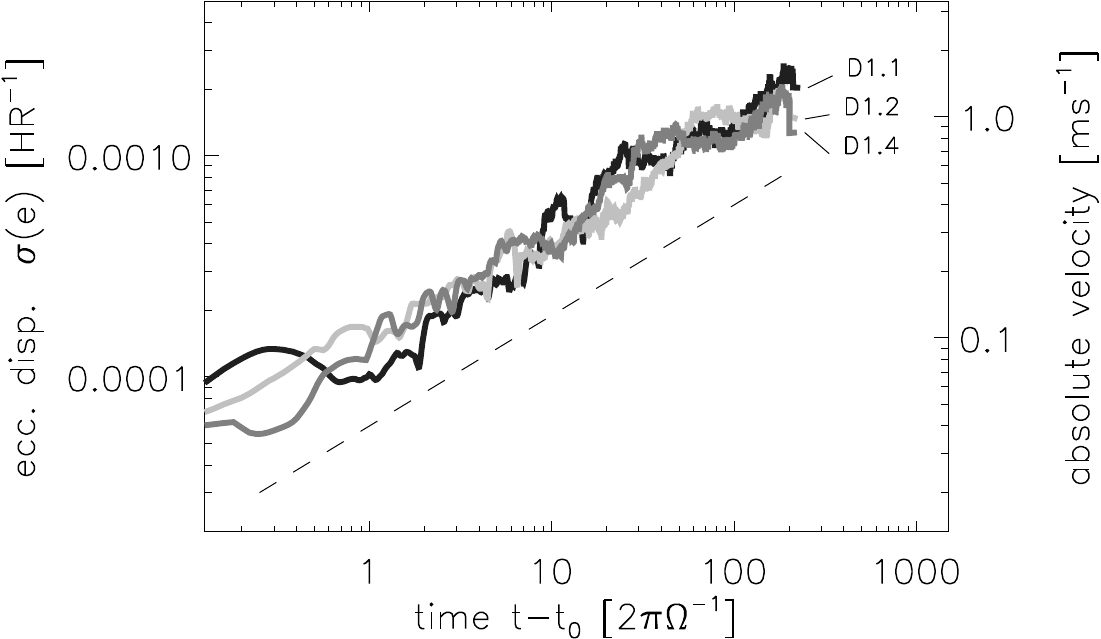}
  \caption{Comparison of the rms random-walk eccentricity $\sigma(e)$, of
    sets of 25 particles in runs D1.1-4. The dashed line shows the
    expected $\sqrt{t}$ behaviour. The second ordinate indicates the
    radial velocity dispersion amongst the particles in absolute
    units. The stirring amplitude is virtually identical in all three
    runs.}
  \label{fig:edisp_cmp}
\end{figure}
Figure~\ref{fig:edisp_cmp} shows that the rms eccentricity curves for
the different models resemble a braided band. Owing to the modest
number of embedded particles, the chance fluctuations are quite
pronounced again. We register an overall spread of the curves of about
a factor of two, which explains the scatter in the fitted amplitudes,
$C_{\sigma}(e)$, describing the random-walk curve
\begin{equation}
  \frac{\sigma(e)}{H/r} = C_{\sigma}(e) \sqrt{t}.
  \label{eqn:e-v-t}
\end{equation}
The values are listed in the last column of
Tab.~\ref{tab:sim_results}, and as a whole agree markedly well with
the result from the earlier run D1 from Paper~II (see Table~3 in that
paper). The striking agreement of the three very different simulations
D1.1-4 in terms of their effect on embedded particles again highlights
the rather stark disagreement with a scaling proportional to the
midplane density fluctuations $\delta\rho$. This will be discussed in
detail in Section~\ref{sec:drho_trq}, and the implications of the
obtained stirring relations for planetesimal evolution will be
discussed in Sect.~\ref{sec:discussion}. Having established the
dependence on the disc mass, we now move on to the effect of the
external net flux permeating our simulation box.


\section{Dependence on external magnetic field}
\label{sec:net_flux}

In the previous discussion, as well as in Papers~I and II, we have
tacitly assumed a fiducial value of about $10\mG$ for the net vertical
magnetic field permeating the accretion disc (see third column in
Tab.~\ref{tab:sim_results}). This particular value was chosen to
produce a turbulent transport coefficient compatible with commonly
accepted T Tauri accretion rates. In a global picture, such a field
naturally arises from local fluctuations -- i.e., it would be hard to
imagine that all the various subsections of a disc manage to retain
zero vertical flux at the same time. \citet{2010ApJ...712.1241S} have
recently verified this notion by dissecting global zero-net-flux
simulations into shearing-box size portions, and found that the
turbulent stresses from the emerging dynamo field match the scaling
derived from local box simulations with an equivalent imposed vertical
net-flux. While their simulations did not include explicit dissipation
terms, there is no obvious reason why this correspondence should be
lost in global models including dead zones
\citep{2010A&A...515A..70D}. Note, however, that from unstratified
local simulations \citet{2000ApJ...530..464F} found a critical
$\Rm_{\rm c}\simeq 10^4$ necessary to sustain MRI in a zero-net-flux
configuration. This threshold was recently confirmed for stratified
discs by \citet{2011arXiv1102.5093O}. Looking at
Fig.~\ref{fig:ppd_strat}, we see that this criterion is only met above
about $3.5\,H$ in our models, leaving very little space for a
self-sustained dynamo mechanism. Because convergence is much harder to
obtain without a net flux \citep[see e.g.][and references
therein]{2010ApJ...713...52D}, we consider only runs that include net
flux vertical fields in this work.

\begin{table*}
\caption{Overview of simulation results including runs from papers I
  \& II. For reference, we list the net vertical field $\rms{B_z}$
  (col. 3) which serves as an input parameter. All turbulent stress
  parameters (cols.~4-6) are normalised by the midplane pressure of
  model D1, i.e., $p_0=1$. Relative fluctuations $\delta\rho/\rho$ are
  midplane values. The width $\sigma(\Gamma_{\!y})$ of the torque
  amplitude is obtained from Fig.~\ref{fig:trq_his}, correlation times
  $\tau_{\rm c}$ are from the ACFs in Fig.~\ref{fig:trq_acf}, and
  $C_{\sigma}(\Delta x)$ and $C_{\sigma}(e)$ are derived from
  Figs.~\ref{fig:xdisp_cmp} and \ref{fig:edisp_cmp},
  respectively. Note that runs NVF$\,$a/b apply different $\tau_{\rm
  diff}$ (cf. Eqn.~(\ref{eq:tau_diff}) in Sect.~\ref{sec:super_box}),
  assuming $k_r=\pi/r$, and $2\pi/r$, respectively. \label{tab:sim_results}}
\begin{tabular}{lcccccccccc}\hline & orbits & net $\rms{B_z}$ 
  & $\rms{\alpha_{\rm Maxw}}$ & $\rms{\alpha_{\rm Reyn}}$ 
  & $\rms{\alpha_{\rm SS}}$ & $\delta\rho/\rho$ & $\sigma(\Gamma_{\!y})$ 
  & $\tauc$ & $C_{\sigma}(\Delta x)$ 
  & $C_{\sigma}(e)$ \\[2pt] 
  & & $[{\rm mG}]$ & & & & & $[\cm^2\s^{\!-2}]$ 
  & $[2\pi\Omega^{-1}]$ & $[H]$ & $[H/r]$ \\[2pt]
  \hline
  A1     & 20-217  & 10.7 & & & 0.0105 & -- & 0.45\ee{8} 
         & 0.30 & 5.21\ee{-3} & 2.68\ee{-3}\\
  D2     & 20-223  & 10.7 & & & 0.0051 & 0.166 & 0.13\ee{8} 
         & 0.27 & 7.25\ee{-4} & 2.50\ee{-4}\\
  B1     & 20-505  & 16.1 & & & $0.05\quad$ & 0.168 & 0.74$^a$\ee{8} 
         & 0.32 & 7.70\ee{-3} & 2.77\ee{-3}\\
  \hline
  D1     & 20-224  & 10.7 & & & 0.0038 
         & 0.103 & 0.06$\ $\ee{8} & 0.29 & 4.72\ee{-4} & 1.54\ee{-4}\\
  D1.1   & 20-230  & 10.7 & 3.17\ee{-3} & 0.64\ee{-3} & 0.0038 
         & 0.091 & 0.054\ee{8}    & 0.30 & 4.56\ee{-4} & 1.67\ee{-4}\\
  D1.2   & 20-222  & 10.7 & 3.22\ee{-3} & 0.59\ee{-3} & 0.0038 
         & 0.058 & 0.052\ee{8}    & 0.29 & 4.36\ee{-4} & 1.13\ee{-4}\\
  D1.4   & 20-221  & 10.7 & 2.75\ee{-3} & 0.55\ee{-3} & 0.0033 
         & 0.038 & 0.052\ee{8}    & 0.22 & 2.98\ee{-4} & 1.27\ee{-4}\\[3pt]
  D1.4b  & 20-75  & 5.37 &  0.19\ee{-3} & 0.86\ee{-3} & 0.0010
         & 0.021 & 0.028\ee{8}    & 0.24 & 1.72\ee{-4} & 6.68\ee{-5}\\
  D1-WF  & 40-220 & 2.68 & 0.46\ee{-4} & 0.22\ee{-3} & 0.0003
         & 0.020 & 0.009\ee{8}    & 0.25 & 0.61\ee{-4} & 3.02\ee{-5} \\
  \hline

  D1-NVF$\,$a & $\;\,$35-50$^b$ & 2.68 & 0.47\ee{-4} & 0.22\ee{-3} & 0.0003
            & 0.029 & 0.009\ee{8} & 0.26 & -- & -- \\                 
            & 60-90   & 5.37 & 0.20\ee{-3} & 0.79\ee{-3} & 0.0010 
            & 0.048 & 0.023\ee{8} & 0.29 & -- & -- \\
            & 100-130 & 10.7 & 0.62\ee{-3} & 2.94\ee{-3} & 0.0036 
            & 0.087 & 0.050\ee{8} & 0.33 & -- & -- \\[3pt]

  D1-NVF$\,$b & 20-50   & 10.7 & 0.62\ee{-3} & 3.27\ee{-3} & 0.0039
            & 0.089 & 0.054\ee{8} & 0.34 & -- & -- \\                 
            & 60-90   & 21.5 & 1.65\ee{-3} & 1.04\ee{-2} & 0.0120
            & 0.130 & 0.062\ee{8} & 0.26 & -- & -- \\
            & 100-130 & 43.0 & 4.89\ee{-3} & 3.43\ee{-2} & 0.0392
            & 0.134 & 0.049\ee{8} & 0.23 & -- & -- \\
  \hline
\end{tabular}\\
\parbox[t]{2\columnwidth}{%
$\qquad^a$corrected for 2D/3D evaluation (cf. fig.~8 in Paper~I),$\quad$
$\ ^b$restricted time interval due to transient ringing at beginning
  of simulation}
\end{table*}

As an alternative to a dynamo-generated disc field, one might envisage
an external stellar field permeating the disc. Given the highly
dynamical star-disc interaction during the T~Tauri phase, strong
magnetic fields are to be expected, and observations find stellar
surface fields of roughly $2-3{\,\rm kG}$
\citep{2007prpl.conf..479B}. Assuming a dipolar field geometry implies
a scaling of $B=B_\star\, (R_\star/r)^3$, and taking $B_\star=3{\,\rm
kG}$, and $R_\star=2\,R_\odot$, we arrive at $\bar{B}_z\sim 2.4\mG$,
and $\sim 20\,\mu{\rm G}$ at $r=1\au$, and $r=5\au$,
respectively. Clearly, the contribution of the central star is very
moderate at $5\au$, but should be considered when looking at models
further in. In the following, we will ignore the actual origin of the
precise value of the imposed field, and simply study how a varying
level of the field strength influences our results.

\subsection{Preliminary considerations}

On theoretical grounds there are two effects which determine the
expected level of turbulence: (i) the unmistakable scaling of the
turbulent stresses with the net flux, and (ii) the extent to which
turbulent activity in the MRI-sustaining regions can penetrate into
the poorly ionised layer near the midplane. As for the first dependence, 
it is now safely established that the saturated state of the MRI depends
critically on the net field \citep{1995ApJ...440..742H,2007ApJ...668L..51P},
and as \citet{2009ApJ...707.1233Y} have shown from unstratified simulations
without dead zones, this should translate directly into the level of
turbulent stirring (cf. their fig. 3). In the following, we will set
out to explore how this scaling is affected by a dead zone and try to
establish the limits of its applicability.

In the case of a net vertical flux (NVF), the most powerful MRI modes
are so-called channel flows, i.e. axisymmetric solutions of the
perturbed magnetic and velocity fields, only depending on the vertical
coordinate $z$. It is a remarkable property of these modes that they
remain exact solutions of the underlying equations far into the
non-linear regime; this property even persists in stratified discs
\citep{2010MNRAS.406..848L}. Given the large-scale nature of these
dominant MRI modes, the question arises whether the flow pattern can
efficiently be broken-down into unordered motion. It is commonly
accepted that this transition into turbulence happens via parasitic
instabilities \citep{1994ApJ...432..213G, 2009ApJ...698L..72P,
2009MNRAS.394..715L} feeding on the unstable MRI modes. A detailed
analysis of how these modes look in the case of a stratified disc can
be found in sect.~4 of \citet{2010MNRAS.406..848L}. The key findings
of this analysis are that (i) the parasitic modes have wavelengths of
about $\lambda_{\rm MRI}/2$, and (ii) their growth rates are
considerably reduced compared to the unstratified case. The latter
finding leads to the conclusion that channel flows are expected to be
persistent, if not dominant, features in stratified discs. This notion
is supported by the `streaks' seen in the density fluctuations
(lower-most panels in Figs.~\ref{fig:spctm_NVFa} and
\ref{fig:spctm_NVFb}), which are remarkably similar to the pinching
effect in fig.~3 of \citet{2010MNRAS.406..848L}. Further evidence for
the strong presence of channel modes comes from the horizontally
averaged radial and azimuthal velocities $\bar{u}_r$, and
$\bar{u}_\phi$ (not shown), which exhibit a remarkable correlation
with $\bar{B}_r$, and $\bar{B}_\phi$, and notably show the opposite
parity compared to their magnetic counterparts. This symmetry is
highly indicative of the solutions derived in
\citet{2010MNRAS.406..848L}, cf. their fig.~1, and a direct
consequence of the underlying set of equations. To conclude this
digression, we note that while the presence of MRI channels was
already mentioned in Paper~II, their appearance is even more
pronounced when going to higher field strength.

\subsection{Turbulence amplitude as a function of net-flux} 
\label{sec:turb_nvf}

\begin{figure}
  \includegraphics[width=\columnwidth]{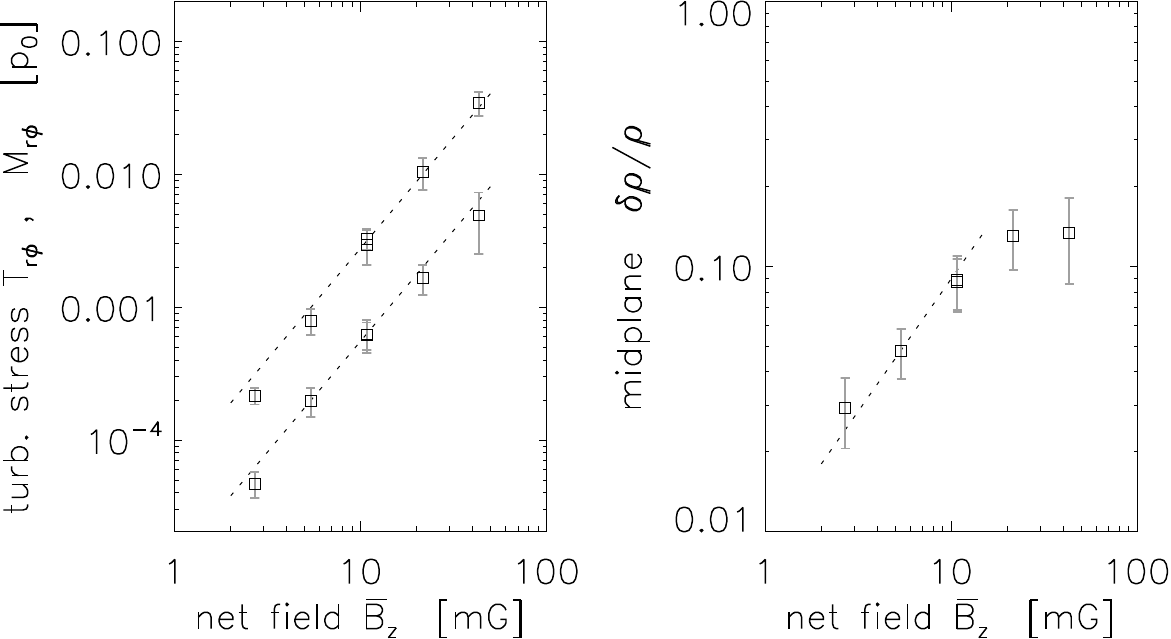}
  \caption{\emph{Left:} Scaling of the dimensionless transport
    coefficients $T_{r\phi}\equiv \rms{\rho\,v_r\delta v_\phi}$, and
    $M_{r\phi}\equiv \rms{-\delta B_r \delta B_\phi}$. The stresses roughly
    scale with $\bar{B}_z$ to the power of $5/3$, as indicated by the
    dotted lines. \emph{Right:} Scaling of the
    relative density fluctuation $\delta\rho/\rho$, measured at the
    disc midplane. Saturation is reached for $\sim20\mG$; below this
    value, the slope is approximately linear.}
  \label{fig:scaling_nvf}
\end{figure}

In the left panel of Fig.~\ref{fig:scaling_nvf}, we plot the time- and
volume-averaged transport coefficients obtained from runs D1-NVFa/b,
covering a range of $\bar{B}_z=2.68$ -- $43\mG$ (i.e. field strengths
spanning a factor of sixteen). We want to point out that this interval
roughly coincides with the practical limits set by the MRI wavelength,
\begin{equation}
  \lambda_{\rm MRI}(z) \equiv
  \frac{2\pi}{\Omega_0}\,\sqrt{15\over16}
  \frac{|\bar{B}_z|}{\sqrt{\rho(z)}}\,,
\end{equation}
of the fastest-growing linear mode. At the weak-field end, the
wavelength is just long enough to be well-resolved, while at the
strong-field end the wavelength approaches the disc thickness (even at
the base of the active layer). These geometric conditions apply to the
case of ideal MHD, and hence are in addition to the requirement $\El >
1$, describing the resistive quenching of the MRI. Note that because
of the z-dependence of the gas density $\rho(z)$, the vertical band
supporting MRI-unstable wave numbers will successively shift towards
the midplane when increasing $\bar{B}_z$. In the limit of strong
fields, this means that the MRI will be increasingly affected by the
$\El>1$ criterion. At the same time, the vertical extend of the
MRI-active band situated between the magnetically-dominated halo and
the resistively-quenched dead zone is expected to shrink, eventually
shutting off the MRI altogether.

Within the permissible range described above, we find that the Maxwell
stress follows the scaling $\propto \lambda_{\rm MRI}^{5/3}$, which is
considerably steeper than the linear dependence proposed by
\citet{2007ApJ...668L..51P} -- see their fig.~2 -- and somewhat
steeper than the $3/2$ dependence suggested by
\citet{2004ApJ...605..321S}. We stress that this does not pose any
inconsistency as the earlier results were obtained for the simplified
case of unstratified MRI. \citet{2007ApJ...668L..51P}, in fact, point
out that their results are likely to be modified when accounting for
the buoyant loss of magnetic energy in the stratified case (this loss
is clearly seen in Figs.~\ref{fig:spctm_NVFa} and \ref{fig:spctm_NVFb}
below).

In passing we note that the canonical mass accretion rate observed in
T Tauri stars is $10^{-8}\Msun\yr^{-1}$, and for disc masses similar
to the minimum mass model this translates into a viscous stress
parameter $\alpha_{\rm ss} \sim 0.01$
\citep{1998ApJ...495..385H}. This value can indeed be achieved in
models below the strong-field limit of the MRI. Setting aside the
significant uncertainties in constraining T~Tauri accretion rates,
this raises the question whether or not a net field of around $20\mG$
(required to generate the observed typical mass accretion rate) can be
sustained by a self-consistent accretion-disc dynamo, or can be
dragged in by the accretion flow that is fed by infall from the parent
molecular cloud.

While the turbulent stress, at very best, shows a weak break around
field strengths of about $10\mG$, the derived midplane density
fluctuation $\delta\rho/\rho$ (see right-hand panel of
Fig.~\ref{fig:scaling_nvf}) shows a clear saturation in the
strong-field limit. To understand this saturation, we again need to
take a look at the vertical structure of the disc.

\subsection{Disc structure at varying net-flux} 

Let us recall that in Section~\ref{sec:disc_mass}, the density
fluctuations at the disc midplane appeared to be determined by the
wave energy available at the dead zone interface defined by the
$\El=1$ condition. For models D1.1-4, the region where the MRI
operated was limited by the column density - i.e. by the disc material
being more efficient in shielding exterior sources of ionisation at
higher mass loading. Because the Elsasser number at the same time
depends on the Alfv{\'e}n speed, and hence the magnetic field
strength, we expect the position of the layer interface to depend on
the external net flux.

\begin{figure*}
  \includegraphics[width=2.1\columnwidth]{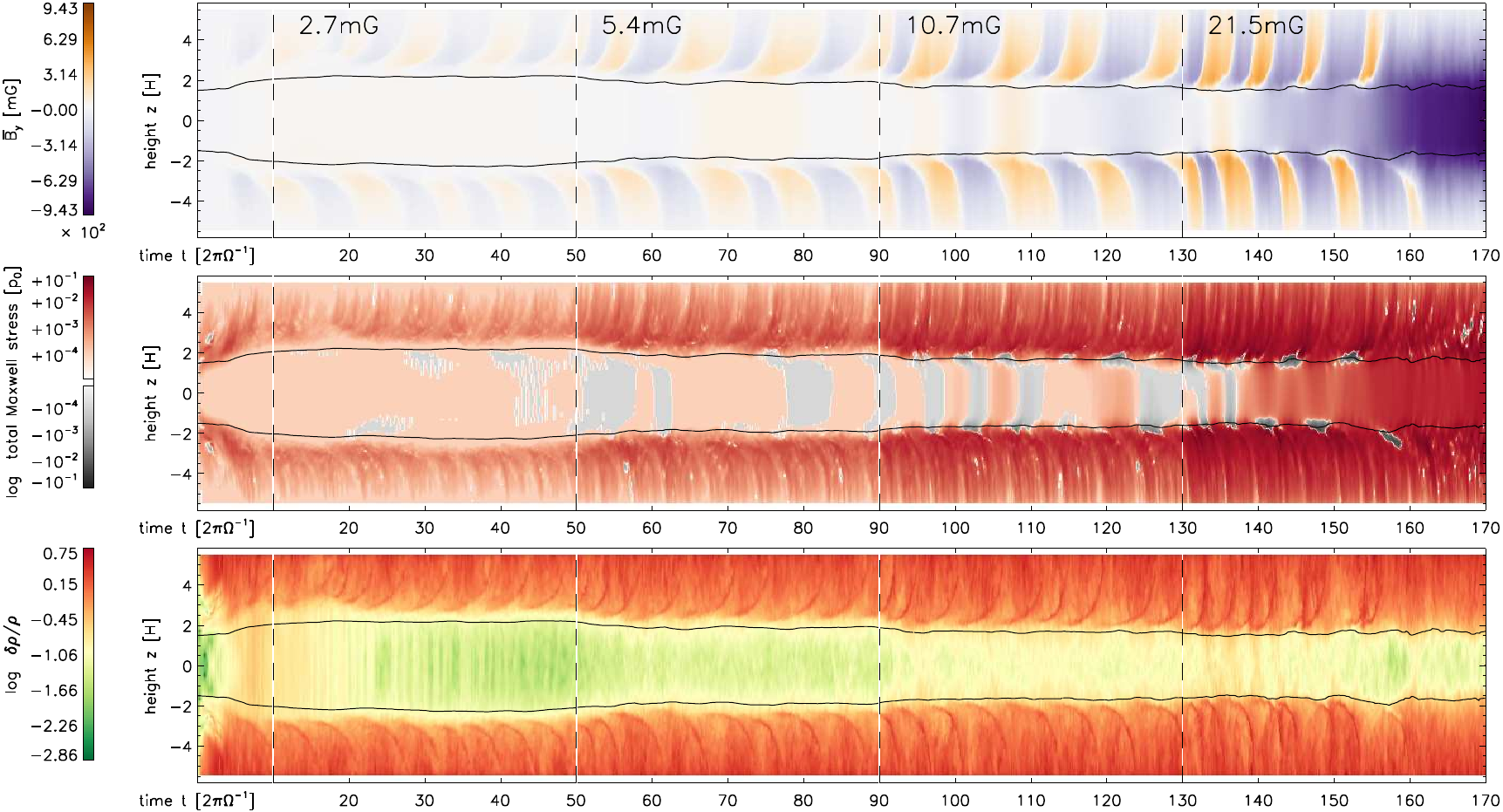}
  \caption{Space-time evolution of the mean toroidal field (upper
    row), total Maxwell stress (middle row), and relative density
    fluctuation (bottom row) for run D1-NVFa. The solid black line
    shows the transition between the MRI-active region and the dead
    zone, approximated by the condition $\El\equiv v_{\rm
    A}^2/(\Omega\eta)=1$. Segmentation indicates intervals with
    different net vertical field strengths of $\bar{B}_z=$ 2.68, 5.37,
    10.73, $21.47\mG$, respectively. The third interval, from 90 to
    130 orbits corresponds to model D1.1, with a net field of
    $10.73\mG$. When obtaining time-averaged quantities, the first 10
    orbits of each interval are disregarded.}
  \label{fig:spctm_NVFa}
\end{figure*}

This dependence is clearly seen in Fig.~\ref{fig:spctm_NVFa}, where we
plot the same quantities as in Fig.~\ref{fig:spctm}, along with the
$\El=1$ interface for the run D1-NVFa. We remind the reader that in
this simulation we increase the net-flux by a factor of two every
forty orbits to examine how the dead zone structure changes (along
with the stochastic forces experienced by embedded planetesimals).
The half-width of the respective dead zones are 2.22$\pm$0.03,
1.89$\pm$0.04, and 1.67$\pm$0.04 for the first three intervals
indicated in Fig.~\ref{fig:spctm_NVFa}. This implies that doubling the
net flux, in this regime, decreases the extent of the dead zone by
about a third of a scale height -- a trend that is equivalent to the
one seen in Sect.~\ref{sec:disc_mass} when reducing the column density
by a factor of two. We note that this direct correspondence, however,
is only valid for weak fields.

To study stronger net vertical fields, we have to overcome an obstacle
which becomes obvious in the last segment of
Fig.~\ref{fig:spctm_NVFa}, where we observe a strong build-up of
toroidal net flux caused by radial field. Diffusing into the midplane
region, this radial flux generates strong azimuthal fields through the
differential rotation. According to \citet{2000ApJ...540..372K}, the
linear growth rate for axisymmetric MRI modes goes asymptotically to
zero at low $\beta_{\rm p}$ if the field has a significant toroidal
component; this likely explains the disappearance of the field
reversals and density streaks (associated with channel modes) as
$\bar{B}_{\phi}$ grows. As discussed in detail in
Sect.~\ref{sec:super_box}, the winding-up of azimuthal field is
artificially enhanced within the shearing box approximation and would
be counter-acted by radial diffusion within a global setup. While we
thwart this artificial build-up via an extra dissipative term
(\ref{eq:diss}) in the induction equation, the adopted time-scale
turned out to be insufficient for net fields exceeding $20\mG$. As a
remedy, we devised an additional simulation run `D1-NVFb' adopting a
more conservative estimate for $\tau_{\rm diff}$, with dissipation
faster by a factor of four. For the new run we chose an initial net
field of $10.7\mG$, which enables us to monitor the effect of the
increased field dissipation term by comparing with the equivalent
segment of run D1-NVFb. As can be verified by inspection of the
respective lines in Tab.~\ref{tab:sim_results}, the derived quantities
agree to within 10\% between the corresponding segments (i.e., orbits
100-130 of NVFa, and orbits 20-50 of NVFb). Moreover, the good
agreement \emph{a posteriori} supports our inclusion of the additional
magnetic diffusion term.

\begin{figure*}
  \includegraphics[width=2.1\columnwidth]{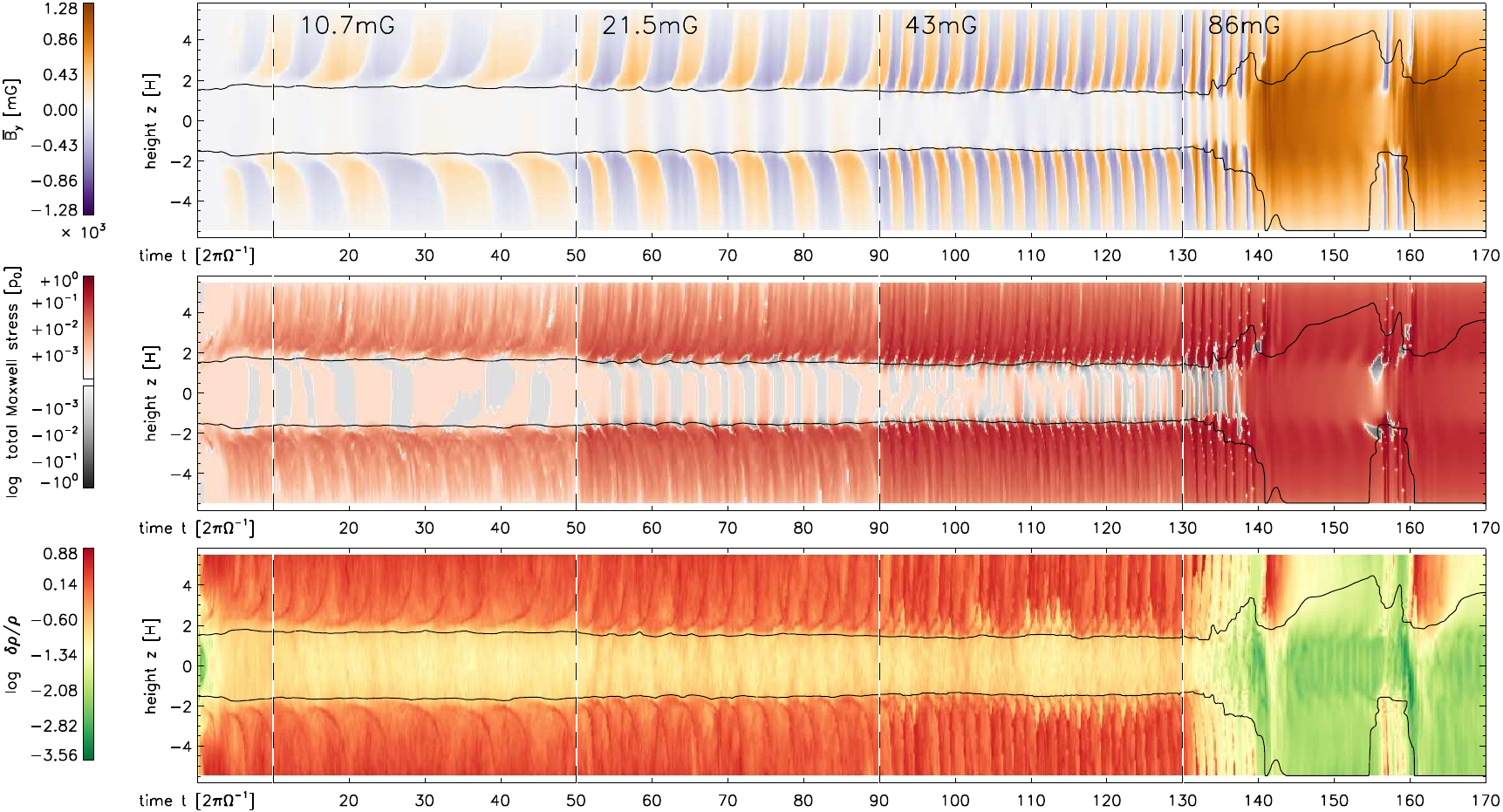}
  \caption{Same as Fig.~\ref{fig:spctm_NVFa}, but for run D1-NVFb with
    a four-times faster super-box-scale dissipation, avoiding the
    artificial field build-up seen at the end of D1-NVFa. Segments
    have different net vertical field strengths as indicated. For the
    strongest field case, we leave the domain in which MRI is possible
    and the disc returns to a laminar state, interrupted only by short
    bursts of activity. Note the different colour bar ranges compared
    to Figure~\ref{fig:spctm_NVFa}.}
  \label{fig:spctm_NVFb}
\end{figure*}

Space-time plots for run D1-NVFb are shown in
Fig.~\ref{fig:spctm_NVFb}, where we see that for fields stronger than
about $40\mG$, MRI cannot be sustained and the disc returns to a
non-turbulent state. Interestingly, in this limit we still observe
sporadic bursts of activity (partly restricted to one ``hemisphere''
of the disc). As above for run D1-NVFa, we infer the half-thickness of
the dead zone by means of the $\El=1$ criterion, and we yield
1.66$\pm$0.03, 1.54$\pm$0.04, and 1.49$\pm$0.10$\,H$, for a net-flux
field $\bar{B}_z=$10.7, 21.5, and 43$\mG$, respectively. We conclude
that by increasing the field strength we narrow the MRI-active region
sandwiched between the dead zone and magnetically-dominated halo, at
the same time approaching a limit in the vertical position of the dead
zone interface. Accordingly, the magnitudes of fluctuations
$\delta\rho$ at the midplane saturate along with the stochastic
torques, as displayed in the right and left panels of
Figs.~\ref{fig:scaling_nvf} and \ref{fig:scaling2_nvf}, respectively.

\subsection{Gravitational torques as function of net-flux}  
\label{sec:trq_nf}

\begin{figure}
  \includegraphics[width=\columnwidth]{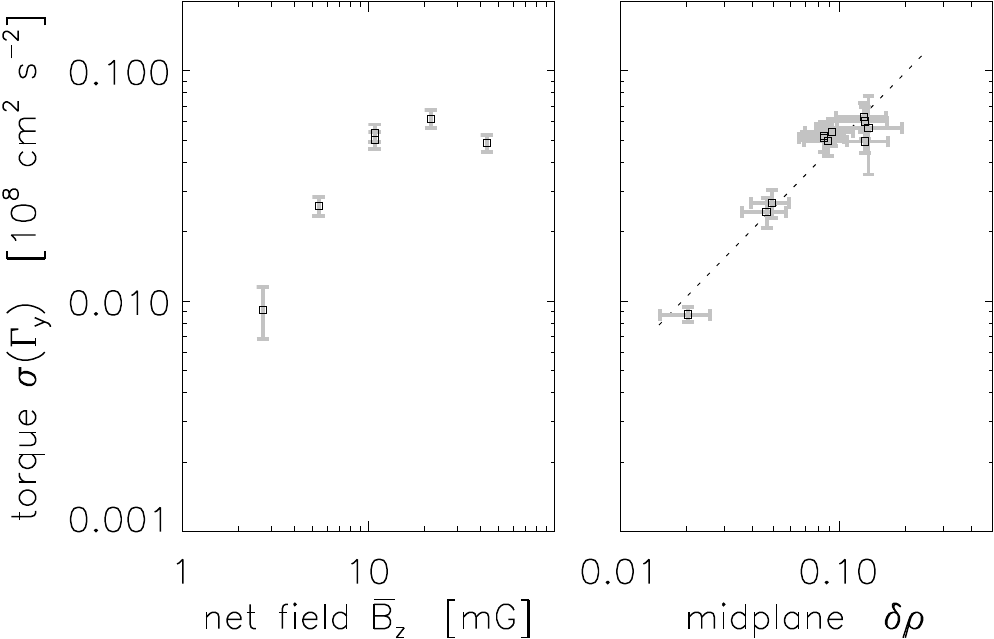}
  \caption{\emph{Left:} Scaling of the rms-torque $\sigma(\Gamma_y)$
    with the net vertical field. \emph{Right:} Scaling with the
    density fluctuation $\delta\rho$ for runs D1-WF (left-most data
    point), and D1-NVFa/b; for the latter, each data point represents
    an interval of $t=15$ orbits. The fitted correlation (dashed line)
    shows a logarithmic slope of $0.97$ in good agreement with the
    expected linear dependence.}
  \label{fig:scaling2_nvf}
\end{figure}

Unlike the runs with varying disc mass discussed earlier in this paper,
we did not evolve the trajectories of embedded particles over time
periods of $\sim 200$ orbits for each value of the magnetic
field. Such an approach would be prohibitively expensive computationally.
Instead, for most models, we monitor the stochastic torques experienced 
by embedded particles by accumulating time series, allowing us to infer
orbital evolution through analysis of the widths of the torque
distributions $\sigma(\Gamma_y)$, and the correlation times $\tauc$
(cf. Sect.~\ref{sec:da-grow} above).  For model D1-WF, corresponding
to a weak field of $2.7\mG$, we evolved the particle trajectories for
200 orbits, and found excellent agreement between the simulation
results and expectations based on the simple diffusion model discussed
in Sect.~\ref{sec:da-grow}, thus justifying our general approach.  The
obtained values can be found in the eighth and ninth column of
Tab.~\ref{tab:sim_results}, respectively.

As expected from the density fluctuations plotted in
Fig.~\ref{fig:scaling_nvf}, the gravitational forcing, for magnetic
field values below $\bar{B}_z\simeq10\mG$, depends strongly on the
level of the applied field (see left-hand panel of
Fig.~\ref{fig:scaling2_nvf}).  For field strengths exceeding $10\mG$,
however, we find saturation of the fluctuating torques, and this is
consistent with the trend found in the density fluctuations in the
preceding section. Moderately stronger fields are already MRI-stable
and models in this regime are likely to lead to a chaotic
time-behaviour of the disc as a whole. Given that we already reached
accretion stresses compatible with limits given by observations of
T~Tauri discs for $\bar{B}_z = 21.5 \mG$, we reckon that the
occurrence of even stronger fields seems unlikely. In this sense, the
obtained torques of around $0.05\times10^8\cm^2\s^{-2}$ appear to be a
firm upper limit for the given dead zone structure. Higher torques are
hence only expected for higher ionising fluxes (cf. model D2 from
Paper~II) or lower dust abundances \citep[not considered here, but
see][]{2009ApJ...703.2152T}, both of which decrease the height of the
dead zone.

To conclude our analysis of the results, we note that the range of
values in $\delta\rho$ and $\sigma(\Gamma_y)$ observed in the three
runs D1-WF, and D1-NVFa/b can be used to infer the correlation between
the two quantities. Recall that for models D1.1-4, we found a somewhat
surprising relation, where a higher value in absolute density
fluctuations $\delta\rho$ would not translate into higher torques. In
the limit of high net flux (and accordingly, high density
fluctuations), this behaviour is compatible with the scatter seen in
the right-hand panel of Fig.~\ref{fig:scaling2_nvf}, where we plot
data points sampled from runs D1-WF, and D1-NVFa/b. For weaker
turbulence (i.e. for lower values of $\bar{B}_z$), however, we
approximately recover the naturally expected linear scaling of
$\sigma(\Gamma_y)$ with the density fluctuations. We conclude that
only looking at the rms-value of $\delta\rho$ in general provides
insufficient characterisation of the underlying gravitational stirring
-- a more sophisticated analysis is required to make quantitative
predictions.


\section{The relation between density wave amplitude and gravitational torques}
\label{sec:drho_trq}

One motivation of this work was to derive intuitive scaling relations
that would allow stochastic torque amplitudes to be expressed as
simple functions of the disc model parameters.  Such functions would
provide straightforward prescriptions for gravitational stirring by
turbulent discs with application to secular evolution models of
planetesimal accretion, et cetera.  The original strategy to achieve
this aim was to break the problem down into three parts, i.e. (a)
predicting the extent of the dead zone and the amplitude of the
turbulence as a function of disc mass and external net flux, (b)
predicting the resulting amplitude of the induced spiral density waves
(characterised by the rms density fluctuation $\delta\rho$ near the
disc midplane), and (c) predicting the associated gravitational torque
experienced by the embedded solids. The third step is particularly
important as the evaluation of gravitational torques within direct
simulations is computationally very expensive when large numbers of
particles are being evolved.

In a recent paper, \citet{2011ApJ...742...65O} have derived simple
predictor functions covering the first two steps outlined above.  With
respect to the third step, it is natural to assume that the resulting
torques scale linearly with the density fluctuation $\delta\rho$, but
contrary to our own expectations, we find that this assumption is not
justified in the presence of a dead zone.  Knowledge of $\delta\rho$
alone is insufficient to construct a simple scaling relation that
describes the gravitational stirring of planetesimals. Indeed the
results presented in Sect.~\ref{sec:disc_mass} are in stark conflict
with this conjecture. In the following, we will elucidate how the
discrepant findings can be reconciled. In our discussion, we are
guided by the simple picture of density waves being excited in the
active layer and then propagating into the dead zone, where they
induce the gravitational stirring experienced by the planetesimals. We
will further refine this picture to include the effect of the
background differential rotation, as appears necessary to explain the
results of our simulations. Finally, we define a set of analytic
formulae that provide a means of estimating stochastic torque
amplitudes as a function of disc parameters.

\subsection{Effects due to background shear} 
\label{sec:shearing_out}

\begin{figure}
  \center\includegraphics[width=0.8\columnwidth]{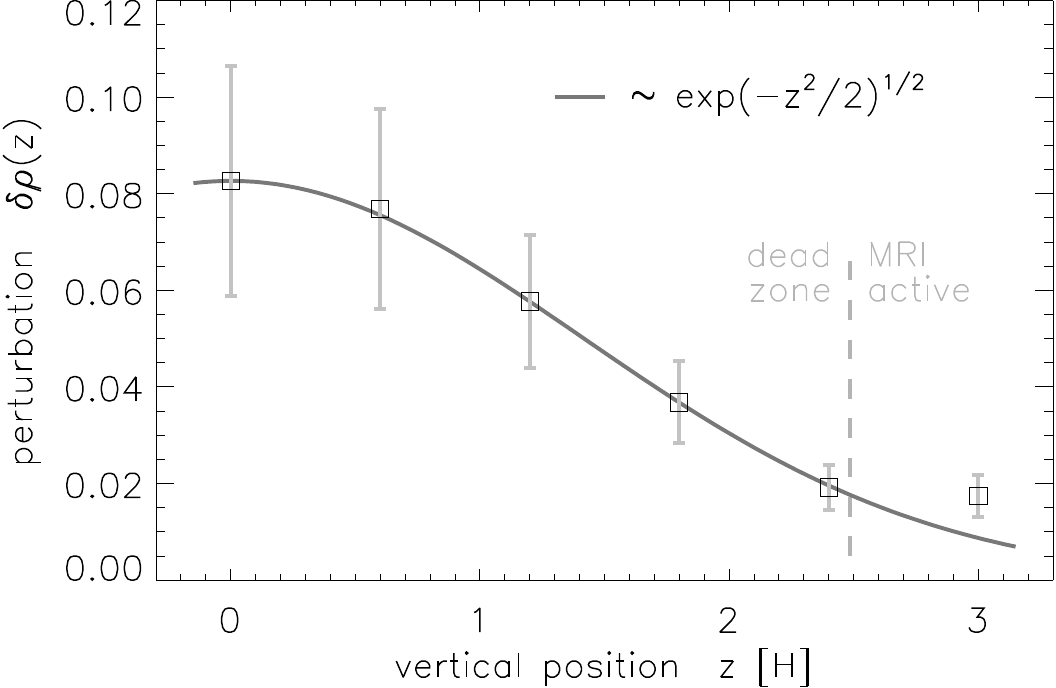}
  \caption{Absolute value of the rms density fluctuation $\delta\rho$
    as a function of the vertical position for a model based on D1.4,
    but with $B_0=5.4\mG$. Within the dead zone, excellent agreement
    with the theoretically predicted scaling based on the exponential
    background density profile is seen.}
  \label{fig:drho_z}
\end{figure}

We recall from the discussion in Sect.~\ref{sec:linear_waves} that
absolute density fluctuations at the disc midplane should obey a
scaling $\delta \rho_{\rm mid} \propto \sqrt{\rho_{\rm mid}}$, which
is indeed observed when comparing models D1.1, D1.2, and D1.4,
respectively.  The agreement in predicting $\delta\rho$ is
independently confirmed when looking at vertical profiles of
$\delta\rho(z)$.  While we only used two particular positions
(i.e. the interface and the midplane) in our previous discussion, we
note that the derived relation should equally hold for any position
within the disc, i.e., $\delta\rho(z) \propto \sqrt{\rho(z)}$. This is
plotted in Fig.~\ref{fig:drho_z}, where error bars indicate the
deviation arising from temporal fluctuations within the adopted time
interval $t=20-85$ orbits. The excellent agreement with the predicted
scaling confirms the assumption based on conservation of wave energy,
and leads to the conclusion that step (b) in the procedure outline
above is justified.

\begin{figure}
  \center\includegraphics[width=\columnwidth]{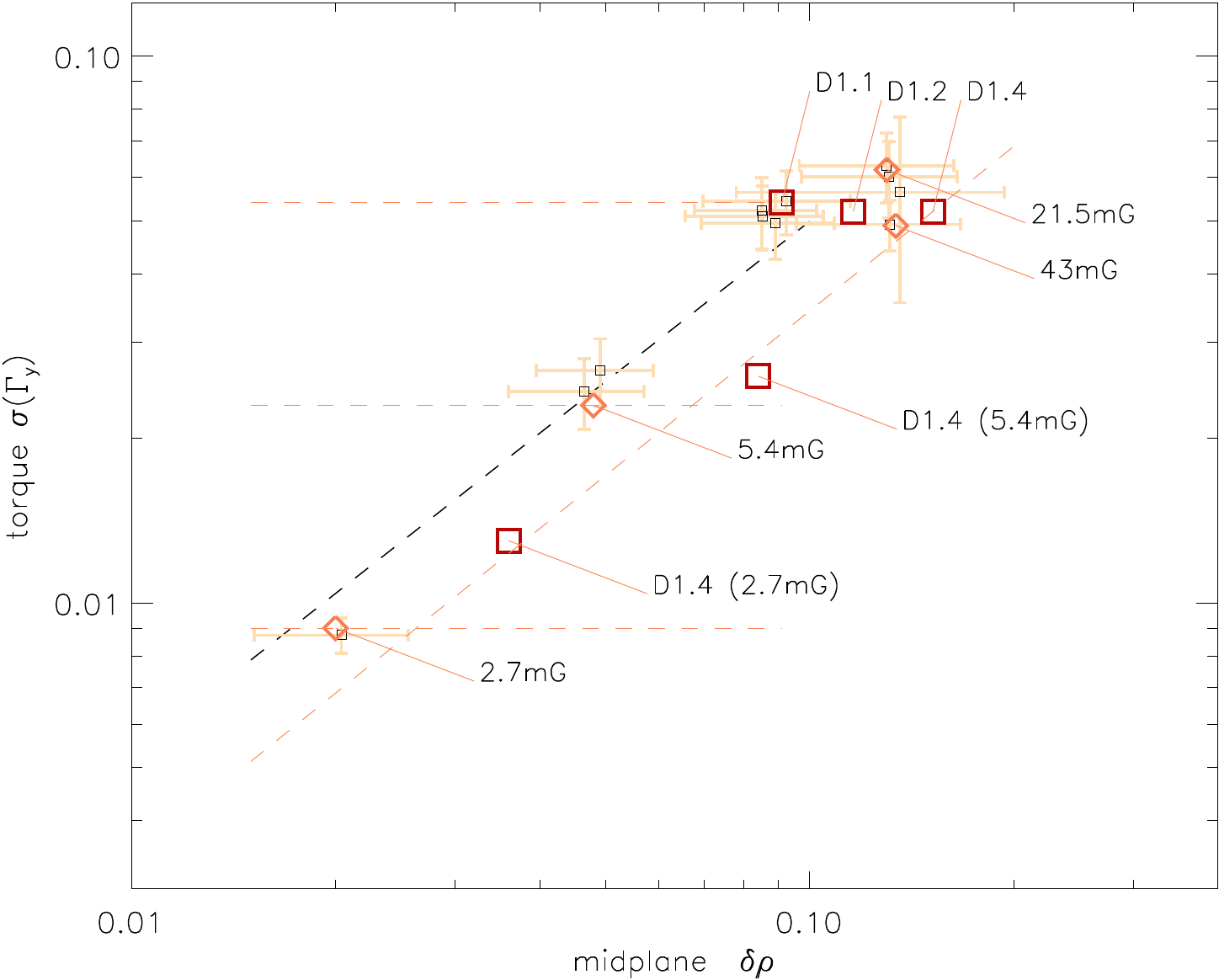}
  \caption{Model overview of the $\delta\rho$ versus torque
    relation. The shaded data points with error bars are taken from
    Fig.~\ref{fig:scaling2_nvf} and complemented by time-averages
    taken over $t=30$ orbits (diamonds). These points represent model
    D1.1 at varying external flux. Additional points (squares) are as
    labelled. Dashed lines indicate a linear and constant scaling,
    respectively.}
  \label{fig:scaling3}
\end{figure}

We now contrast the assumption of linear dependence between the
absolute density fluctuation $\delta\rho$ and its associated torque
$\sigma(\Gamma_y)$ -- to be used in step (c) -- with the results
obtained from a comprehensive set of numerical
simulations. Figure~\ref{fig:scaling3} compiles all available data
points in the $\delta\rho$ versus $\sigma(\Gamma_y)$ plane, including
the ones previously shown in the right-hand panel of
Fig.~\ref{fig:scaling2_nvf} (i.e. for the dependence on the vertical
net flux). We have also included additional runs in
Fig.~\ref{fig:scaling3}, performed to provide improved coverage of the
$\sigma(\Gamma_y)$ -- $\delta\rho$ plane. These runs were for the
heavier disc D1.4 but with external magnetic field values $\bar{B}_z =
5.4$ and $2.7\mG$ (labelled in the figure). Clearly the scatter seen
in the plot cannot be reconciled with a single linear relation between
$\delta\rho$ and the resulting torque. Let us hence focus on a few
particular sub-sets of models first.

Evidently, the torque remains constant\footnote{Note that a similar
pattern (of a too-low torque at increased disc mass) holds when
comparing D1.1 and D1.4 at weaker fields (i.e., $5.4\mG$, and
$2.7\mG$), albeit the effect is seen strongest at $10.7\mG$.} when
going from model D1.1 to D1.2, and D1.4 as discussed already in
Section~\ref{sec:disc_mass}. In contrast, the torques for model
D1-NVFa/b (with varying net-flux at constant disc mass), are
consistent overall with a linear scaling -- although the torques
exhibit a somewhat steeper than linear fall-off as a function of
$\delta \rho$ for $\delta\rho\simlt 0.1$.  The same trend holds for
the additional set of runs based on model D1.4 but with varying net
flux (cf. the red squares on the lower right in
Fig.~\ref{fig:scaling3}) -- here very little deviation from the linear
scaling is seen. The most apparent discrepancy arises between model
D1.1 and a four-times heavier disc at half the vertical net-flux,
labelled `D1.4 ($5.4\mG$)'. While these models agree in the midplane
value of $\delta\rho$ to within 10\%, they differ in their effective
torques by a factor of two. This clearly demonstrates that there has
to exist a significant \emph{qualitative} difference in the density
structure between the two models. The obvious difference between the
strong-field/low-mass model and the weak-field/high-mass model is the
width of the resulting dead zone. As we will see in the following
discussion, this observation provides the key to understanding the
apparent discrepancy.

\begin{figure}
  \begin{center}
    \includegraphics[width=0.85\columnwidth]{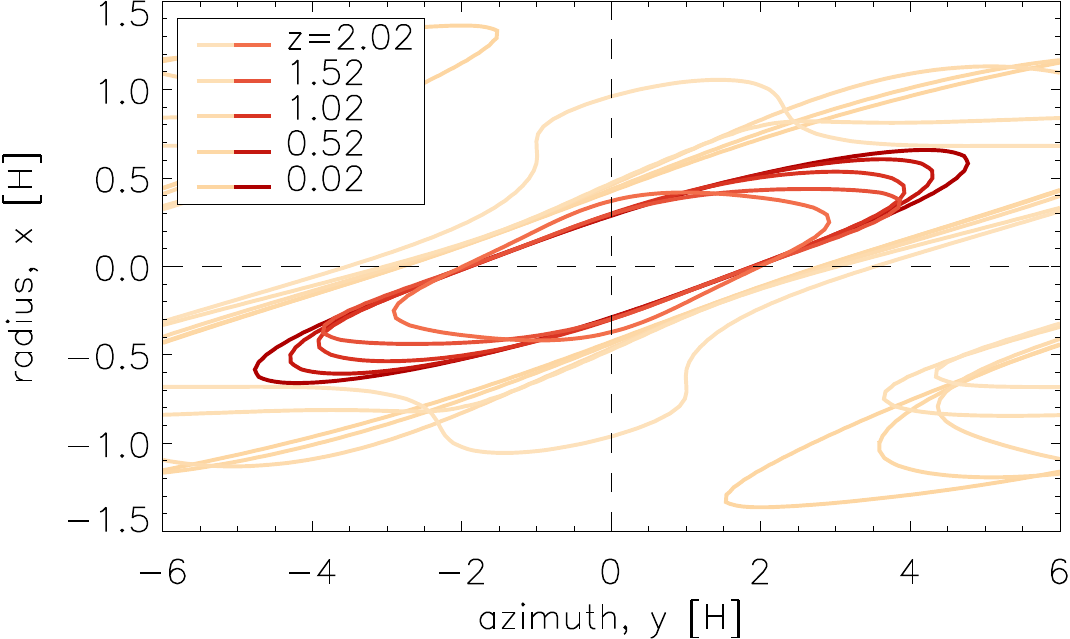}\\[4ex]
    \includegraphics[width=0.85\columnwidth]{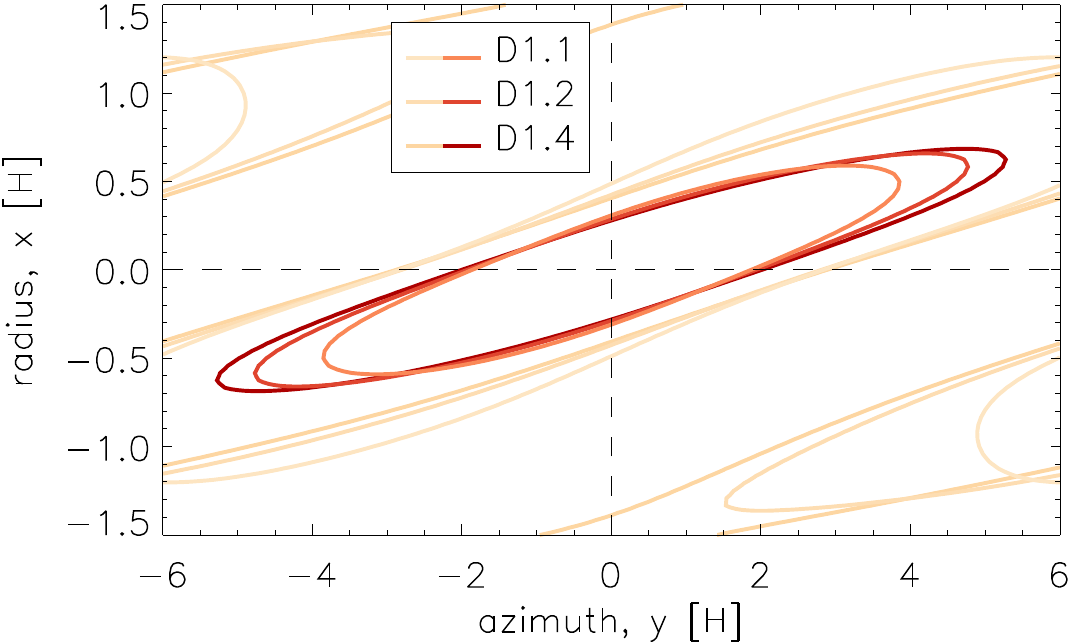}
  \end{center}
  \caption{\emph{Top:} Contours of the two-point density correlation
    function for model D1.2 at various vertical positions in the disc.
    The five dark contours show where the functions fall below 1/e of
    their peak values; light contours represent the first
    zero-crossings. \emph{Bottom:} Comparison of the two-point ACF
    computed within $z=\pm H/2$ of the disc midplane, for models D1.1-4
    (also cf. figs.~3 \& 6 in Paper~I), showing the same trend as
    above, but now due to the different width of the respective dead
    zones.}
  \label{fig:rho_acf}
\end{figure}

\subsubsection{Two-point density correlation} 

In the following, we build on the picture of linear sound waves
transporting density perturbations into the dead zone region as
outlined in Sect.~\ref{sec:linear_waves}. We conjecture that the
shearing-out of these radially and vertically propagating spiral
density waves affects the torque distribution inferred near the disc
midplane. This picture gains support from looking at the two-point
correlation of the gas density, which we compute in horizontal slabs
of vertical extent $z=\pm H/2$, and which we average over $t=20-220$
orbits. The two-point ACF of the density field serves as a qualitative
indicator for the degree of coherence in the density structures and
was used in Paper~I to compare local and global models of fully active
MRI. To trace the characteristic evolution of the propagating density
waves, we compute two-point ACFs at different vertical positions in
the disc, starting from the dead/active zone interface and moving
towards the midplane in steps of $H/2$. This is exemplified for model
D1.2 in the upper panel of Fig.~\ref{fig:rho_acf}, but the observed
trend holds for all the studied models. The plot clearly shows that
density features, on average, are more sheared-out towards the
midplane.

We further conjecture that the stretching arises because the density
waves become increasingly sheared out by the background differential
rotation as they propagate from the vertical position where they are
created down toward the midplane. To estimate the amount of
stretching, we assume a vertical propagation speed of $c_{\rm
s}=H\,\Omega$, resulting in a propagation time $\Delta t=2\,H/c_{\rm
s}=2\Omega^{-1}$ for a dead zone half-width of two scale
heights. Assuming Keplerian shear with $q=3/2$, and evaluating at a
position $x\simeq 0.5\,H$ away from the coordinate origin, this
amounts to an azimuthal displacement of $\Delta y=q\,\Omega\, x\,
\Delta t=3/4\,H\Omega\times 2\Omega^{-1}=1.5\,H$, which is in decent
agreement with the level of stretching seen in
Fig.~\ref{fig:rho_acf}. As a further step, we checked that the
characteristic pitch angle for density waves observed at the dead zone
interface is the same for all the models.

With this knowledge at hand, we can now attempt to reconcile the
results from Sect.~\ref{sec:disc_mass}. Given the significant widening
of the dead zone when increasing the disc mass, we expect more
sheared-out ACFs for model D1.4 than for model D1.2, and D1.1,
respectively. This is because the travel time for waves to reach the
planetesimals situated near the disc midplane directly depends on the
width of the dead zone. The suggested trend is clearly seen in the
lower panel of Fig.~\ref{fig:rho_acf}, where we plot the midplane ACFs
for the three models.

The shapes of the ACFs are characteristic of trailing spiral waves,
and the degree of correlation in the azimuthal direction shows a clear
trend with disc mass. This, in turn, supports the notion that the
stronger fluctuations $\delta\rho$ in the heavier disc models are
compensated by the more elongated aspect ratio in the density
structures. It is well known that the winding-up of the spiral
structures due to differential rotation leads to a linear increase in
the radial wave-number $k_x\propto (3\Omega/2)\,t\,k_y$ in time. This
process goes along with a phase-folding of the azimuthal
structures. The enhanced azimuthal symmetry of increasingly sheared
out density waves, as viewed by embedded planetesimals, leads to a
reduction in the average amplitude of the stochastic torque due to
partial cancellation of the induced azimuthal gravitational
acceleration.

\subsubsection{Modified scaling relation} 

\begin{figure}
  \center\includegraphics[width=\columnwidth]{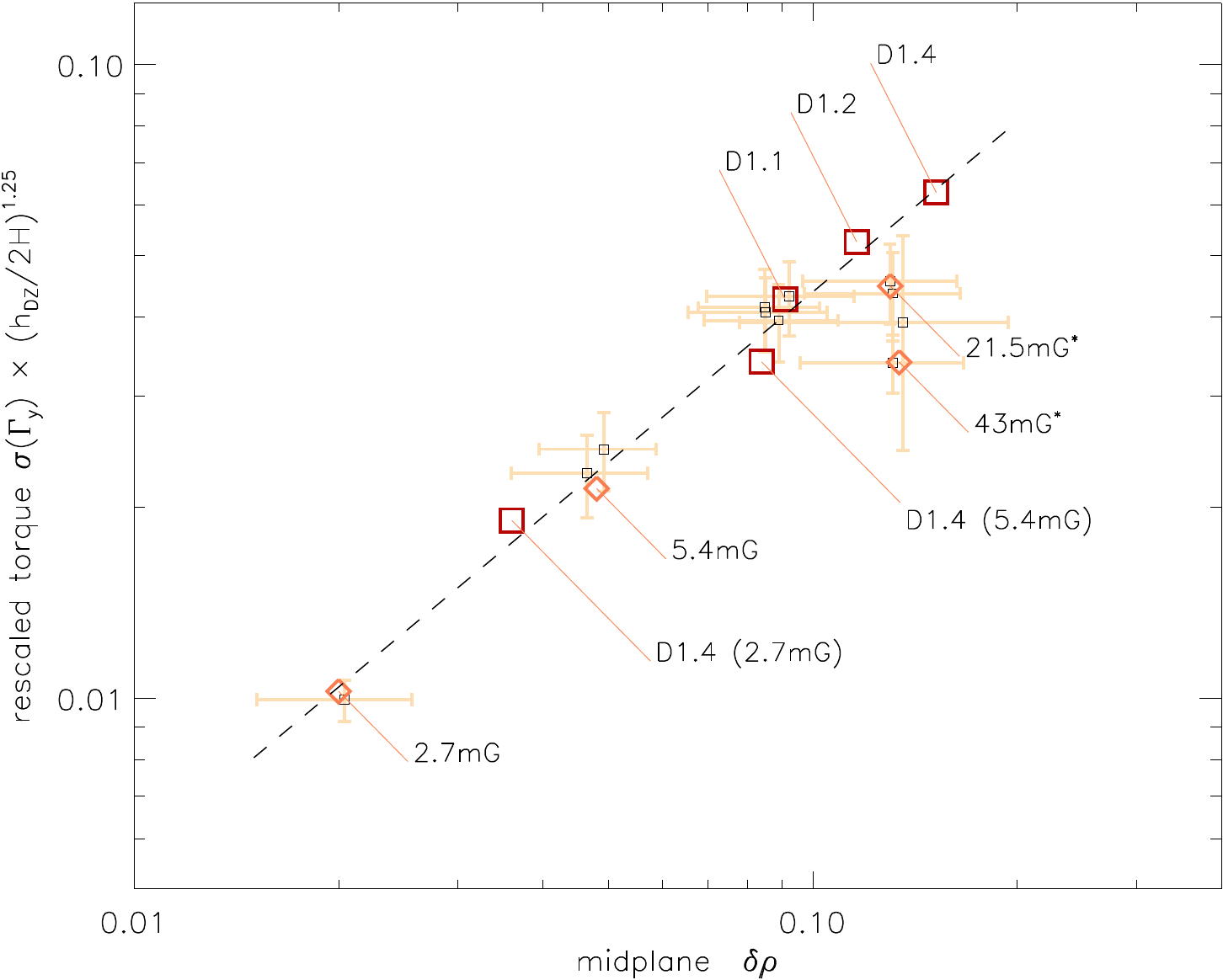}
  \caption{Same as Figure~\ref{fig:scaling3}, but with the torques
    rescaled by $(h_{\rm DZ}/2H)^p$ to compensate for the attenuation
    of the measured torques by the stretching-out of the density
    field. Excluding the strong-field models (see text), we obtain a
    best-fit correction parameter of $p\simeq 5/4$.}
  \label{fig:scaling4}
\end{figure}

So far, our reasoning has been largely qualitative. In this section we
aim to provide more quantitative evidence of how the simple linear
relation between density fluctuations and gravitational torques might
have to be modified. Having identified the distance away from the dead
zone interface as the relevant parameter determining the pitch angle
of the density field (and hence the effective gravitational force), it
appears natural to use the dead zone half-width $h_{\rm DZ}$ as a
parameter to empirically correct our prediction for the resulting
torques. This is done in Fig.~\ref{fig:scaling4}, where we boost the
measured torques by a factor $(h_{\rm DZ}/2H)^p$, with $p$ to be
determined. The rescaling is done to restore, as accurately as
possible, a linear dependence $\sigma(\Gamma_y) \propto
\delta\rho$. The particular choice of the fiducial width $2H$ implies
that we pivot around model D1.2, which has this dead zone width and
hence remains unaffected. We point out that the rescaled torque
amplitude does not have any physical implication but merely serves as
a benchmark. In practice, one can use the inverse scaling factor to
correct torque estimates based on predicted values of $\delta\rho$ for
the attenuation caused by the stretching.  By varying $p$ and
inferring deviations from a power-law, we find a best-fit index of
$p\simeq 5/4$, which does a reasonable job in collapsing the scatter
plot to a single line in log-log space. As indicated by stars in
Fig.~\ref{fig:scaling4}, we excluded the data points with field
strengths higher than the nominal value of $10.7\mG$. This has been
done because in this limit the dead zone width does not change any
more, and the stirring amplitude saturates accordingly. We note,
moreover, that the resulting slope is somewhat smaller than the
expected linear relation. Given the simplicity of the approach,
however, we regard this a satisfactory result. As can be seen in
Fig.~\ref{fig:scaling4}, the rescaling mostly affects models with
different column densities, and the models D1.1, D1.2, and D1.4 now
show the expected linear scaling with $\delta\rho$. This also affects
the heavy disc models with weaker fields (D1.4 [2.7$\mG$], and D1.4
[5.4$\mG$]), which have a wider dead zone. We remark that the
rescaling does not destroy the good fit already seen in these models,
but at the expense of a power-law index deviating from one. Moreover
the slight deviation from the linear relation within the data points
belonging to different time intervals in models D1-NVFa/b
(diamond-shaped points \& D1.1) is now much smaller. Overall we regard
the procedure as reasonably successful in correcting for the
attenuation of the torque amplitudes suffered because the density
waves shear out as they propagate towards the disc midplane.  We hence
suggest to compensate torque estimates based on $\delta\rho_{\rm mid}$
by an extra factor $(h_{\rm DZ}/2H)^{-1.25}$ with respect to a
fiducial model with $h_{\rm DZ}=2H$. Because heavier discs have wider
dead zones and hence weaker effective torques (compared to the
original scaling based on $\delta\rho_{\rm mid}$ alone), in practice,
this leads to more favourable conditions for planetesimal accretion in
heavier discs, where gas-drag damping of the induced velocity
dispersion is enhanced.

\subsection{A simple stochastic torque prescription} 
\label{sec:torque_scaling}

Having derived the additional correction factor for the stochastic
torque amplitude as a function of $h_{\rm DZ}$, we are now in the
position to provide a simple analytic fitting formula that
encapsulates all the data points of the studied set of models. We here
focus on providing rms torque amplitudes $\Gamma_y$ and correlation
times $\tauc$.  Owing to saturation effects, we limit the scope of
applicability to magnetic field strengths of $\bar{B}_z \simlt
10.7\mG$.  We also point out that the derived scalings are only
strictly valid for the chosen ionisation model, representative of a a
protosolar nebula at a distance of $a=5\au$, and for a range of column
densities around a few times the MMSN.

As we have seen in sections \ref{sec:disc_mass} and
\ref{sec:net_flux}, the width of the dead zone changed by about a
third of a scale-height when either doubling the disc mass, or halving
the magnetic net-flux, respectively. More precisely, we obtain a
half-width
\begin{eqnarray}
    h_{\rm DZ} = 1.66\,H 
               & - & \Delta h_{\rm B}\, 
                     \frac{\ln(\bar{B}_z/10.7\mG)}{\ln\,2} \nonumber\\
               & + & \Delta h_{\Sigma}\,
                     \frac{\ln(\Sigma/135\g\cm^{-2})}{\ln\,2}
    \label{eq:hdz}
\end{eqnarray}
relative to model D1, and with coefficients $\Delta h_{\Sigma} =
0.33\,H$, and
\begin{equation}
    \Delta h_{\rm B} = 0.28\,H 
                     - 0.04\,H\,\frac{\ln(\Sigma/135\g\cm^{-2})}{\ln\,2}\,,
    \label{eq:dhb}
\end{equation}
i.e., accounting for a slightly weaker dependence of $h_{\rm DZ}$ on
the net magnetic field when going to heavier disc models.

The scaling of the rms midplane density fluctuation $\delta\rho$ with
disc mass was derived in Sect.~\ref{sec:linear_waves}. Based on the
empirical finding of $\delta\rho_{\rm mid}/\rho_{\rm mid} \propto
\rho_{\rm mid}^{-0.63}$, we here use a power-law index of $0.37$,
i.e., deviating slightly from the analytically predicted square-root
dependence. For the scaling with the vertical net flux, the assumption
of a simple linear dependence between $\delta\rho$ and $\bar{B}_z$
produced equally good results for the models based on D1.1, and D1.4,
respectively. Overall we thus arrive at a predictor function for the
rms stochastic torque amplitude of
\begin{equation}
    \Gamma_y =  \Gamma_{y,{\rm D1}}\,
                \left( \frac{\Sigma}{135\g\cm^{-2}} \right)^{0.37}
                \left( \frac{\bar{B}_z}{10.7\mG} \right)\,
                \left( \frac{h_{\rm DZ}}{1.66\,H} \right)^{-1.25},
\end{equation}
relative to the fiducial model D1 with $\Gamma_{y,{\rm D1}} =
0.054\times10^8 \cm^2\s^{-2}$, and with $h_{\rm DZ}$ as given by
(\ref{eq:hdz}), and (\ref{eq:dhb}).  This formula -- together with a
canonical value of $\tauc\simeq 0.3$ orbits -- completes our simple
prescription for including stochastic torques in generic planetesimal
evolution models.


\section{Long-term evolution}
\label{sec:discussion}

In Sects.~\ref{sec:e-growth} and \ref{sec:trq_nf} we discussed the
rate at which the velocity dispersion of planetesimals is excited by
gravitational interaction with turbulent density fluctuations in the
disc midplane. The driving of the velocity dispersion is counteracted
by the combined effects of gas drag and collisional damping
\citep{2008ApJ...686.1292I}, leading to a well defined equilibrium
value for the rms random velocity. We now calculate the approximate
magnitude of this equilibrium velocity dispersion and compare it with
estimates for the catastrophic disruption thresholds of planetesimals
composed of strong or weak materials.

\subsection{Planetesimal equilibrium velocity dispersions} 

Following closely the analysis presented in sect.~5.1 of Paper~II, we
estimate the equilibrium velocity dispersion, $v_{\rm disp}$, for
planetesimals of different sizes, and compare it with the catastrophic
disruption thresholds for strong and weak aggregates to determine
under which conditions collisions between (equal-sized) planetesimals
will lead to growth rather than destruction.

\begin{figure*}
  \begin{center}
      \includegraphics[height=0.8\columnwidth]{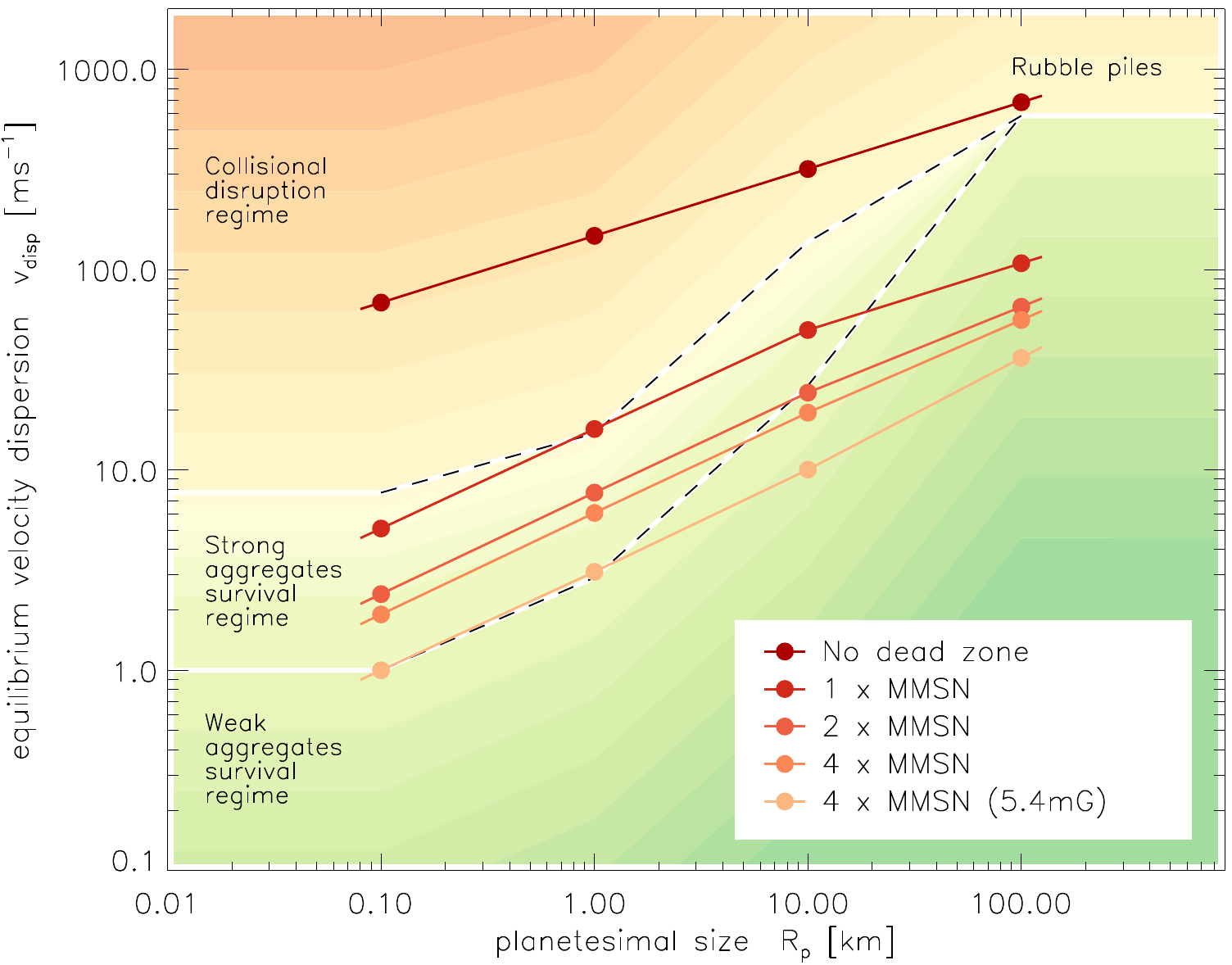}\hspace{4ex}
      \includegraphics[height=0.8\columnwidth]{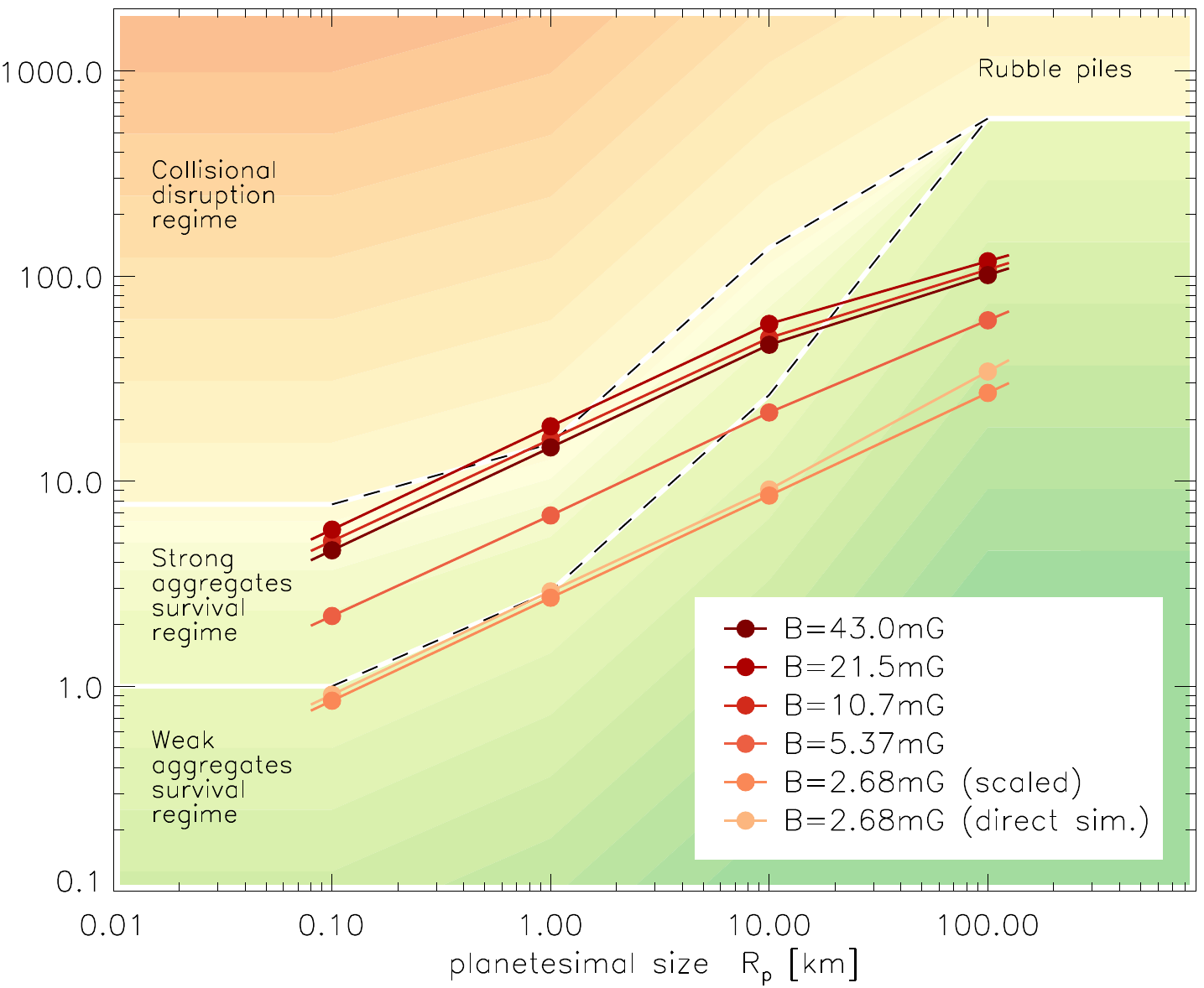}
  \end{center}
  \caption{\emph{Left:} Variation of the equilibrium velocity
    dispersion as a function of planetesimal size and the disc
    mass. Increasing the disc mass increases the midplane density and
    planetesimal collision probability, thereby increasing gas drag
    and collisional damping.  \emph{Right:} Variation of equilibrium
    velocity dispersion as function of planetesimal size and magnetic
    field strength.}
  \label{fig:vdisp}
\end{figure*}

Balancing the rate of eccentricity excitation obtained in simulations
with the gas drag damping rate, we obtain the following expression for
the resulting velocity dispersion (see Paper~II for details)
\begin{equation}
  v_{\rm disp}=\left[ \frac{4 C^2_{\sigma}(e) h^2 R_{\rm p} 
      \rho_{\rm p} v_{\rm k}^2}{
      10^9 C_{\rm D} \rho} \right]^{1/3}\cm\s^{-1}\,.
  \label{eq:vdisp-drag}
\end{equation}

Similarly, balancing the rate of eccentricity growth with the mutual
collisional damping rate for planetesimals (assuming all solids in the
disc have grown into planetesimals of a given size, $R_{\rm p}$, with
a mean density $\rho_{\rm p}=2\g\cm^{-3}$, and the coefficient of
restitution $\simeq 0$) gives the expression
\begin{equation}
  v_{\rm disp} = \sqrt{ \frac{4 R_{\rm p} \rho_{\rm p} 
      C_{\sigma}^2(e) h^2 v_{\rm k}^2}{
      10^9 \Sigma_{\rm p} \Omega_{\rm k}}}\;\cm\s^{-1}\,.
  \label{eq:vdisp-coll}
\end{equation}
The catastrophic disruption thresholds for strong (basalt) and weak
aggregates are given by \citet{2009ApJ...691L.133S}, and are discussed
in detail in Paper~II in the context of planetesimal stirring in
turbulent discs. We plot the catastrophic collision velocities in
Fig.~\ref{fig:vdisp} as a function of planetesimal radius (solid black
lines). We assume that the largest $100\km$ sized bodies are
rubble-piles rather than monolithic bodies, and so adopt the
appropriate disruption velocity in this limit. In this figure we also
plot the equilibrium velocity dispersions for planetesimals of
different size from the different simulations. The plotted values
correspond to the smaller of those calculated using
eqns.~(\ref{eq:vdisp-drag}) and (\ref{eq:vdisp-coll}). The left panel
shows results for different disc masses with a fixed external magnetic
field strength of $10.7 \mG$. The right panel compares results for
different magnetic field strengths in a disc with fixed mass
(approximately equal to the MMSN model).

The left panel of Fig.~\ref{fig:vdisp} shows that planetesimals with
radii between 0.1 and $1\km$ have equilibrium velocity dispersions
that lie between the catastrophic disruption thresholds for weak and
strong materials when embedded in discs with masses between 1 and 4
times the MMSN. Collision outcomes for these bodies will depend
sensitively on the material properties. $10\km$ bodies in discs with 2
or 4 times the MMSN mass will have velocity dispersions that are
marginally lower than the catastrophic disruption thresholds for even
weak materials. $100\km$-sized rubble piles are clearly safe from
disruption in all models. The decreasing equilibrium velocity with
increasing disc mass arises because the gas drag scales with the
midplane density, the mutual collision rate depends on the number
density of planetesimals, and the strength of the stochastic stirring
was found to be essentially independent of the disc mass for these
models.

Examining the right panel of Fig.~\ref{fig:vdisp}, we see that
reducing the value of the magnetic field leads to a reduction in the
stirring of planetesimals. Decreasing the field strength increases the
size of the dead zone and reduces the level of turbulence in the
active layers.  We remind the reader that for most simulations
conducted with varying magnetic field strength, the orbital
trajectories of particles were not evolved under the influence of disc
forces, but instead we monitored the time varying torque due to the
turbulent fluctuations acting on the particle swarm.  Estimates for
the values of $C_{\sigma}(e)$ used in eqns.~(\ref{eq:vdisp-drag}) and
(\ref{eq:vdisp-coll}) to calculate the equilibrium velocity
dispersions were obtained by assuming a linear scaling between the rms
fluctuating torque and $C_{\sigma}(e)$, using the run with $\bar{B}_z
=10\mG$ to provide the normalisation. The validity of this approach is
confirmed by the run with $\bar{B}_z=2.68\mG$ for which particle
trajectories were evolved under the influence of disc forces, and for
which both the predicted and simulated values of $C_{\sigma}(e)$ were
found to be in very good agreement, as shown in the right panel of
Figure~\ref{fig:vdisp}. For this run, which contains the weakest field
considered, the equilibrium values of $v_{\rm disp}$ fall beneath the
catastrophic values appropriate for weak aggregates for all
planetesimal sizes, suggesting that all bodies will be safe from
collisional destruction in such a disc.  For stronger fields, however,
the values typically lie between the thresholds for strong and weak
materials. $100\km$-size rubble-piles are again safe from collisional
break-up in all models. We note that the predicted mass accretion rate
for the disc model with $\bar{B}_z=2.68\mG$ is rather low, being
${\dot M} \simeq 5\times 10^{-9}\Msun\yr^{-1}$.  This is a factor of
two lower than the canonical value of $10^{-8}\Msun\yr^{-1}$ observed
for T Tauri stars \citep{2004AJ....128..805S}. The scaling we obtain
with field strength and disc mass, however, suggest that a slightly
stronger field can generate a larger accretion rate in the active
zone, and in a more massive disc than the MMSN the velocity dispersion
will remain lower than the catastrophic values for weak
materials. This expectation is supported by the simulation D1.4b
described in Tab.~\ref{tab:sim_results}, which used the heaviest disc
model D1.4 and an external magnetic field $\bar{B}_z=5.4
\mG$. Turbulent stresses induce a mass accretion rate $\sim 2.7 \times
10^{-8}\Msun\yr^{-1}$, and the values of $v_{\rm disp}$ are
illustrated by the lowest set of points in the left panel of
Fig.~\ref{fig:vdisp}.  It is clear that disc models can be constructed
that satisfy requirements on disc accretion rates and allow even the
weakest planetesimals to avoid collisional disruption.  It is advisable
not to over-interpret the results of our study, however, due to the
simplifying assumptions that have gone into our calculations. Suffice
it to say that we have been able to produce disc models with
reasonable accretion rates within which the dead zone provides a
safe-haven against catastrophic disruption of planetesimals of all
sizes even when they are composed of essentially strength-less
materials.

\subsection{Implications for planet formation} 

We now summarise the key results from our studies of planetesimals
embedded in turbulent discs, focusing on their relevance for planet
formation theory. We do this by first providing a brief review of our
findings from Papers~I and II and then discuss how these are affected
in view of the new results obtained in this paper.

In Paper~I, we considered the dynamical evolution of small particles
(pebbles and boulders) whose interaction with the gas was dominated by
gas drag, in addition to larger bodies (planetesimals) whose
interaction is dominated by gravitational interaction with turbulent
density fluctuations. Particles tightly coupled via gas drag have
sizes $\lesssim 1\m$ at $5\au$, and were shown to rapidly achieve a
velocity dispersion equal to the gas turbulent velocity. At the
midplane in a dead zone this velocity is typically 10-$30\ms$,
possibly slowing protoplanet growth in models that rely on accretion
of small particles by planetesimals and embryos
\citep[e.g.][]{2004AJ....128.1348R,2010MNRAS.404..475J,2010A&A...520A..43O}.

We now focus on the evolution of larger bodies whose dynamical
evolution is determined by gravitational interaction with turbulent
density features. Papers~I and II demonstrate that a model of
planetesimal formation based on gradual binary sticking of smaller
particles cannot operate in a fully turbulent disc due to rapid growth
of the velocity dispersion above catastrophic disruption thresholds
for bodies of size $\lesssim 10\km$.  Models that invoke rapid
formation of large (i.e., $\sim100\km$) planetesimals may operate
\citep[e.g.][]{2007Natur.448.1022J,2008ApJ...687.1432C}, but runaway
growth to form larger bodies is not possible because the velocity
dispersion quickly rises to the surface-escape velocity of these
objects. Forming planets within the disc lifetime appears to be
difficult in a fully turbulent disc without a dead zone, unless a
model involving gravitational instability is invoked
\citep{1998ApJ...503..923B}.

Results from Paper~II (and model D1.1 in this work) show that
stochastic forcing of planetesimals is $\sim 20$ times weaker for a
minimum-mass disc with a nominal dead zone than in a disc without a
dead zone. Equilibrium velocity dispersions for $100\m$ - $10\km$
bodies lie between disruption thresholds for weak and strong materials
\citep{2009ApJ...691L.133S}, being between a few metres and a few tens
of metres per second (cf. sect.~5.3 in Paper~II for details).  In
principle, planetesimals that form via gradual accumulation can
survive mutual collisions in the nominal model if composed of
moderate-strength materials. To form in the first place, however, they
need to overcome barriers to growth occurring at mm and metre sizes
\citep{2008A&A...480..859B, 2010A&A...513A..57Z}.  $10\km$-sized
bodies forming gradually will not experience runaway growth to form
larger oligarchs as their velocity dispersion reaches $\sim 10\ms$
within $\sim 10^5\yr$. Because the dispersion remains below the
destruction threshold, further growth can proceed; eventually delayed
runaway growth may occur for resulting bodies with $\sim 100\km$ radii
\citep[also see][]{2010Icar..210..507O}.

Rapid formation of large planetesimals with sizes $\gtrsim 100\km$
through turbulent concentration of chondrules
\citep{2008ApJ...687.1432C} or the streaming instability
\citep{2007Natur.448.1022J} in the nominal dead zone results in
runaway growth because of slow turbulent stirring of the velocity
dispersion. Self-stirring by planetesimals determines the dispersion
in this case rather than turbulence. We note that all dead zone models
predict that $100\km$ rubble-piles are safe against turbulence-induced
collisional destruction.

Increasing the disc mass by a factor of two or four does not
dramatically change the above picture, as demonstrated by models D1.2
and D1.4 in Sect.~\ref{sec:disc_mass}. These models result in smaller
equilibrium dispersion velocities for larger disc masses because of
the constant stirring strength and increased eccentricity
damping. Increasing the external magnetic field strength above $10.7
\mG$ increases stochastic forcing and $v_{\rm disp}$, but saturation
is attained when the external field approaches values too strong for
the MRI to operate in the active zone.  Reducing the field strength
weakens stochastic forcing by increasing the dead zone width and
reducing activity levels in the turbulent layers.  A model with
$\bar{B}_z=2.7\mG$ results in $v_{\rm disp}$ values below the
catastrophic disruption threshold for the weakest aggregates. This
important result demonstrates that weak $100\m$ - $100\km$
planetesimals can survive collisional destruction in a reasonable dead
zone model, albeit one whose mass accretion rate is a factor of two
below the canonical mass accretion rate in T Tauri stars (but still
within the observed range of values).  Model D1.4b demonstrates,
however, that a heavier disc with a slightly stronger field can
generate a larger mass accretion rate and similar velocity dispersion
(see Tab.~\ref{tab:sim_results} and Fig.~\ref{fig:vdisp}).  Weaker
stochastic forcing leads to slower growth of the velocity dispersion
toward the escape velocity of $10\km$-sized bodies. The time for the
velocity dispersion to reach $10\ms$ in the weak magnetic field model
is $\sim 3\Myr$, so runaway growth of $10\km$-sized bodies can occur
in this disc.


\section{Conclusions}
\label{sec:conclusions}

This is the third paper in a series that examines the dynamics of
planetesimals embedded in turbulent protoplanetary discs. In the
present work we have used local disc models situated at $5\au$ that
include a simple equilibrium gas-grain chemistry to examine how
changing the disc column density and the external magnetic field
strength modifies the size of the dead zone, and the level of
gravitational stirring of particles located at the disc midplane. We
consider column densities that lie approximately between one and four
times the minimum mass solar nebula values, and external magnetic
field strengths between $2.68$ - $43\mG$. Our main conclusions can be
summarised as follows:

\begin{enumerate}

\item Increasing the disc mass by a factor of two or four above the
  nominal value for the minimum-mass model leads to the unexpected
  result that the strength of particle stirring is independent of disc
  mass. Scaling relations based on a simple picture of linear waves
  being excited in the active layer and propagating into the dead zone
  predict that absolute density fluctuations at the midplane will
  scale with the square-root of the column density. This scaling is
  observed in the simulations, but gravitational stirring is found to
  be insensitive to the larger absolute density fluctuations.

\item Analysis of two-point correlation functions for the midplane
  density fields between discs with different dead zone heights show
  that typical density structures in discs with larger dead zones are
  more elongated in the azimuthal direction. Density waves propagating
  from the active layer into the dead zone are more sheared out when
  they reach the midplane in larger dead zones, resulting in reduced
  stochastic gravitational forces.

\item We suggest this phenomenon explains the insensitivity of
  planetesimal stirring to increasing disc mass, as is demonstrated by
  a scaling-law for the stochastic torques that depends on the dead
  zone vertical extent, and which accounts reasonably for our
  simulation results.

\item Decrementing the external magnetic field strength increases the
  size of the dead zone and decreases the level of activity in the
  turbulent layers. This, in turn, leads to reduced absolute density
  fluctuations in the disc midplane and corresponding reductions in
  the level of gravitational stirring experienced by embedded
  planetesimals.

\item Moderate increments in the external magnetic field strength
  reduce the size of the dead zone, and increase the level of activity
  in the turbulent layers. This, at first, results in an increased
  stirring of planetesimals, but quenching occurs when the position of
  the dead zone interface reaches saturation. This is because the
  amount to which the active layers can constrict the dead zone is
  limited (via the $\El=1$ criterion) by the steep rise in
  $\eta(z)$. For our model, torque fluctuations saturate for
  $\bar{B}_z \simgt 20\mG$, while turbulent stresses grow up to net
  fields of $\simeq 40\mG$, where the MRI eventually shuts off.

\item A disc hosting a modest external field (of $\bar{B}_z = 2.7 -
  5.4\mG$) leads to equilibrium velocity dispersions for $100\m$ to
  $10\km$ planetesimals below the catastrophic disruption threshold
  for even the weakest aggregates. This demonstrates that reasonable
  disc models can be constructed in which the dead zone provides a
  safe-haven for weak planetesimals against collisional destruction.

\end{enumerate}

There remain a number of caveats that we have not yet addressed, and
which may modify the detailed conclusions drawn in this paper. These
include adoption of a simple equation of state for the disc, and a
mono-disperse size distribution for planetesimals when estimating
collisional damping rates.  The most important omissions are the Hall
effect and ambipolar diffusion from the induction equation, which may
affect the size of the dead zone and the level of turbulent activity
in the active zones \citep{2011arXiv1103.3562W}. Nonetheless, we
expect that the main result of our study remains robust, namely that
dead zones lead to much reduced stochastic forcing of planetesimal
random velocities, and provide potentially benign locations for
planetary formation.


\section*{Acknowledgements}

We thank the anonymous referee for her/his detailed report. This work
used the \NIII code developed by Udo Ziegler at the Leibniz Institute
for Astrophysics (AIP). All computations were performed on the QMUL
HPC facility, purchased under the SRIF initiative. N.J.T. was
supported by the Jet Propulsion Laboratory, California Institute of
Technology, the NASA Origins and Outer Planets programs, and the
Alexander von Humboldt Foundation.


\appendix


\section{Numerical convergence check} %
\label{sec:resol}

\begin{figure}
  \center\includegraphics[width=0.9\columnwidth]{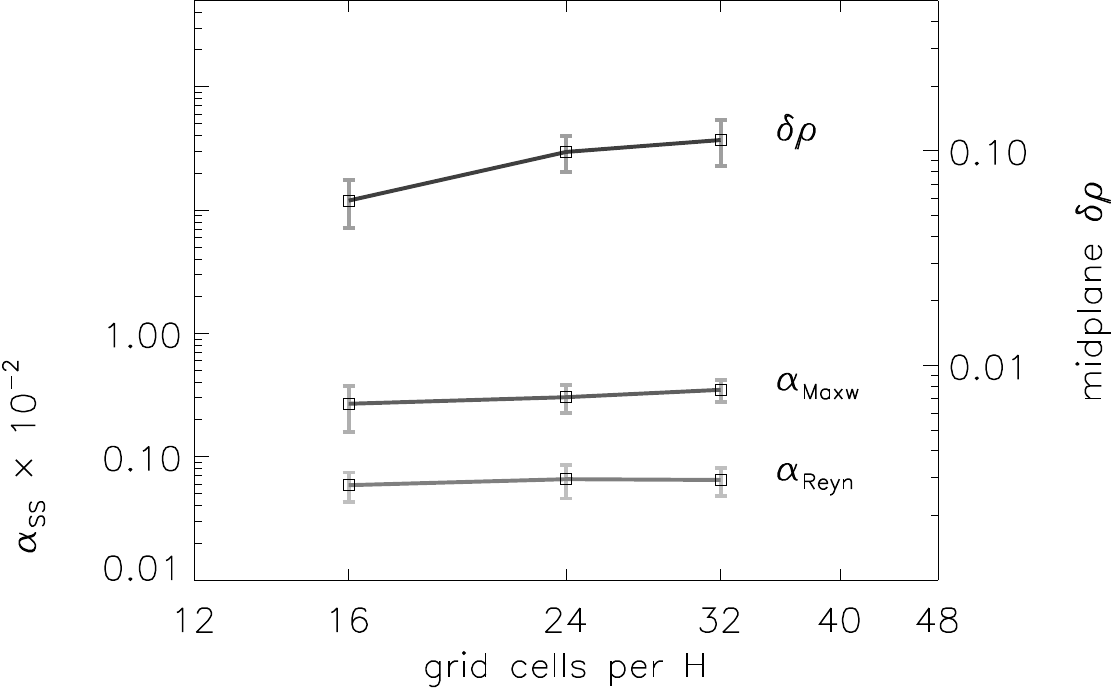}
  \caption{Dependence of key quantities on the numerical
    resolution. Error bars indicate fluctuations within the used time
    interval of $t=20-50$ orbits. All results presented in this
    work adopt a value of 24 grid cells per $H$.}
  \label{fig:resol}
\end{figure}

To verify that our simulations are numerically converged, we performed
two additional runs identical to model D1.1, but with the numerical
resolution changed to 16, and 32 grid cells per $H$, respectively.
Owing to constraints in computational cost, we were not able to
evaluate gravitational torques for the higher-resolved run but have to
rely on the hydrodynamic quantities as proxies instead. The dependence
of key quantities defining the turbulent state of the fluid are shown
in Figure~\ref{fig:resol}.

As was briefly discussed in Sect.~\ref{sec:turb_nvf}, the numerical
requirements for resolving the development of the MRI in the presence
of a net vertical field are safely established by linear
theory. Accordingly, we see very little change in the turbulent
stresses $\alpha_{\rm Maxw}$, and $\alpha_{\rm Reyn}$ when increasing
the numerical resolution beyond the basic requirement. However, from
our study of the box-size dependence in Paper~I, we recall that
convergence in these quantities does not necessarily guarantee that
secondary effects are equally well resolved. In particular, this
applied to the convergence (with box size) in the amplitude
$\delta\rho$ of the excited of spiral density waves. Because of the
predicted non-linear steepening of the analytic wave profile
\citep{2011arXiv1109.2907H}, we suspect that spurious numerical
dissipation might have an influence on the lifetime of wave fronts and
hence on the overall amplitude of density perturbations.

This effect is seen in the dependence of $\delta\rho$ on the numerical
resolution as shown by the upper line in Figure~\ref{fig:resol}.
There $\delta\rho$ changes by almost 70\% when increasing the number
of points from 16 to 24 per $H$, i.e., for a 50\% increase in linear
resolution. In contrast, when going from 24 to 32 cells (corresponding
to an additional 33\% increase in resolution), the change in
$\delta\rho$ remains below 15\%, such that the discrepancy between the
two resolutions is consistent within the error bars. Put into
proportion with the level of accuracy aimed for, this indicates that
our standard resolution of $24/H$ is reasonably converged and will
allow to predict turbulent density fluctuations to within about $\pm
25\%$.


\section{Modifications to the induction equation} %
\label{sec:ind_mod}

Let us first focus on the time step limitations imposed by the high
value of the diffusivity coefficient $\eta$ near the disc midplane
(cf. Fig.~\ref{fig:ppd_strat}). Given the need to evolve the
simulations over several hundred dynamical time scales to follow the
evolution of the embedded planetesimals, using these profiles is
prohibitive. By inspecting earlier simulations and test runs, however,
we have determined that there exists no significant small-scale
structure in the regions of high diffusivity. This is hardly
surprising as the diffusion term is very efficient at removing
features in the magnetic field topology. We have found that imposing a
cap to the diffusivity, $\eta$, corresponding to a value of $\Rm=3$,
does not significantly change the observed field topology, while at
the same time permitting a much larger computational time
step. However, as we will see below, additional care has to be taken
to make this approach viable.

\subsection{Evolution of mean fields} 

Owing to the periodic nature of the shearing box approximation, the
only field structures that remain in the dead zone region are vertical
variations in the radial and azimuthal field. Magnetic fields are
essentially uniform in the horizontal directions. The vertical
variations are given by the horizontally-averaged mean values
$\bar{B}_x(z)$, and $\bar{B}_y(z)$. Let us now examine how these are
affected by diffusion. Note that the induction equation
\begin{eqnarray}
  \partial_t \B -\nabla\tms(\V\tms\B -\eta \nabla\tms\B) & \!=\! &  0\,,
  \nonumber\\      \nabla\cdt\B & \!=\! &  0\,,
  \label{eq:ind}
\end{eqnarray}
is linear in both $\B$, and $\eta$, and can be decomposed
accordingly. For reasons that will become obvious in a moment, we
split the induction equation into two parts, reflecting the evolution
of the horizontally averaged mean fields and their fluctuations
\citep[cf.][]{2010MNRAS.405...41G}. Because the geometric averaging
operation trivially commutes with the differential operators in
(\ref{eq:ind}), this decomposition is exact. Focusing on the evolution
of the mean fields, we now write down the diffusive part of the
related induction equation:
\begin{eqnarray}
  \partial_t\,\bar{B}_i(z) & = & 
  \dots+\partial_z\,\left[\ \eta(z)\ \partial_z\bar{B}_i(z)\ \right]\,,\qquad 
  i \in \left\{ x,y \right\} \,.
  \label{eq:mean}
\end{eqnarray}
Introducing a cap on the value of $\eta$ (and hence removing its
$z$-dependence beyond a certain depth within the dead zone) may
significantly alter the character of (\ref{eq:mean}), and we
illustrate this by means of a simple example: Note that for constant
$\eta$, a field varying linearly with $z$ is not subject to any
further diffusion as $\eta \partial_z^2\bar{B}_i(z)=0$ in this
case. In contrast, any ``bump'' in $\eta$ will make $\eta(z)\
\partial_z\bar{B}_i(z)$ vary in $z$ even if $\partial_z\bar{B}_i(z)$
is constant. The diffusion term will then act to flatten a given
linear ramp, removing any residual gradients in $\bar{B}(z)$. Looking
at the strong gradients $\partial_z\,\eta(z)$ seen in
Fig.~\ref{fig:ppd_strat}, this implies a potentially efficient means
to redistribute field. Ultimately, a weak gradient in $\bar{B}$ can
lead to strong diffusion in the presence of a large gradient in
$\eta$.

Because it is a one-dimensional equation, the overhead of solving
(\ref{eq:mean}) is computationally inexpensive, even under stringent
time step constraints. Consequently, we split (\ref{eq:ind}) into four
parts by separating both $\B=\mB+\B'$ and $\eta = \eta_{>3}({\bf x},t)
+ \eta_{<3}(z,t)$.  Here subscripts refer to the resulting magnetic
Reynolds number, $\Rm$, as indicated by a dashed line in
Fig.~\ref{fig:ppd_strat}; also note the function arguments indicating
that the highly diffusive low-$\Rm$ part of $\eta$, (i.e. $\eta_{<3}$)
is only applied to the mean-field equation. In summary, this approach
allows us to retain the full $\eta$~profile in (\ref{eq:mean}), while
using the capped $\eta$~profile in the expensive three-dimensional
equation. We finally remark that retaining the full $\eta$ in the mean
fields will likely enhance the leakage of field into the dead zone, as
discussed above.

\subsection{Accounting for super-box scale effects} 
\label{sec:super_box}

Neglect of the large-scale radial structure makes the shearing-box a
poor set-up for global magnetic field evolution. This deficiency
becomes most apparent within the dead zone, as we have discussed
briefly above.  Accounting for the global disc topology,
\citet{2008ApJ...679L.131T} derived a criterion for the evolution of
large-scale magnetic fields within dead zones. Owing to the
limitations imposed by the neglect of curvature terms, shearing boxes
cannot reproduce the related threshold, which applies in global
cylindrical coordinates. This is because the divergence of radial
field lines is not included, so $B_x$ can be uniform where $B_r$ would
decline as $1/r$.  Omitting radial gradients in $B_r$ contributes to
the uniform shear-generated $B_{\phi}$ and the weak resistive
diffusion.

Moreover, in a global disc model, the Keplerian rotation profile
$\Omega(r)\propto r^{-3/2}$ leads to a radial variation of the
shearing time scale. In contrast, a single orbital frequency
$\Omega_0$ applies throughout the local box, such that a uniform
radial field stretches out into a uniform toroidal field. With the
toroidal field uniform, the resistive term $\eta\,\nabla^2\,B_{\phi}$
in the induction equation toroidal component is zero, preventing the
diffusion of the shear-generated fields. As a result, local
simulations of dead zones can exhibit the growth of large toroidal
flux \citep[cf.][]{2011ApJ...732L..30H}, which may spuriously
feed-back into the MRI-active regions.

One possibility to account for the global field evolution would be to
remove toroidal flux at a rate consistent with the solution of the
global induction equation. For example, if one assumed a power-law
radial dependence for the dead zone toroidal field $B_\phi\propto
r^p$, and a radially uniform resistivity, the axisymmetric cylindrical
induction equation has resistive term $\partial_t B_\phi =
\eta\,B_\phi\,(p^2-1)\,r^{-2}$. Because of the uncertainty related to
the parameter $p$, we chose an even simpler approach and estimate the
time scale on which diffusion redistributes azimuthal flux as
\begin{equation}
  \tau_{\rm diff} = \frac{1}{k_r^2\eta}\,,\qquad
  \textrm{where we chose}\ k_r \equiv \pi/r\,.
  \label{eq:tau_diff}
\end{equation}
Noting that a Fourier mode of wave number $k_r$ would exponentially
decay according to $\exp(-t/\tau_{\rm diff})$, we remove toroidal (and
radial) flux via an additional source term
\begin{equation}
  \partial_t\,\bar{B}_i(z) = \dots - \bar{B}_i(z)/\tau_{\rm diff}\,
  \label{eq:diss}
\end{equation}
in the one-dimensional part of the induction equation. As we will find
when discussing the results, this choice of the global decay time
scale retains some weak fields within the midplane \citep[also
cf. fig.~3 in][]{2011ApJ...732L..30H}, indicating that we are not
significantly over-damping the system. As can be seen in the upper
panel of fig.~5 of Paper~II, model D1 is affected by relatively strong
fields diffusing into the midplane region. To make the comparison with
models D1.2, and D1.4 as straight-forward as possible, we decided to
re-run this simulation including the diffusive sink term in
Eqn.~(\ref{eq:diss}), giving the new version the label `D1.1'.


\bsp

\label{lastpage}

\end{document}